\documentclass[reprint,amsmath, superscriptaddress, amssymb, aps, longbibliography, prl]{revtex4-1}

\usepackage{graphicx}
\usepackage{dcolumn}
\usepackage{bm}

\usepackage[T1]{fontenc}
\usepackage[utf8]{inputenc}
\usepackage{physics}

\usepackage{xcolor}
\newcommand\hl[1]{%
  \bgroup
  \hskip0pt\color{blue}%
  #1%
  \egroup
}
\DeclareMathAlphabet{\pazocal}{OMS}{zplm}{m}{n}
\newcommand{\Tb}{\pazocal{T}}

\newcommand{\Da}{\mathcal{D}}

\begin{document}

\title{Ultrafast phonon-driven charge transfer in van der Waals heterostructures}

\author{Giuseppe Meneghini}
\email{giuseppe.meneghini@physik.uni-marburg.de}
\affiliation{%
 Department of Physics, Philipps University of Marburg, 35037 Marburg, Germany}%

\author{Samuel Brem}

\affiliation{%
 Department of Physics, Philipps University of Marburg, 35037 Marburg, Germany}%

\author{Ermin Malic}
\affiliation{%
 Department of Physics, Philipps University of Marburg, 35037 Marburg, Germany}
 \affiliation{%
 Department of Physics, Chalmers University of Technology, 41258 Göteborg, Sweden
}%

\date{\today}

\begin{abstract}
Van der Waals heterostructures built by vertically stacked transition metal dichalcogenides (TMDs) exhibit a rich energy landscape including interlayer and intervalley excitons. Recent experiments demonstrated an ultrafast charge transfer in TMD heterostructures. However, the nature of the charge transfer process has remained elusive. Based on a microscopic and material-realistic exciton theory, we reveal that phonon-mediated scattering via strongly hybridized intervalley excitons governs the charge transfer process that occurs on a sub-100fs timescale. We track the time-, momentum-, and energy-resolved relaxation dynamics of optically excited excitons and determine the temperature- and stacking-dependent charge transfer time for different TMD bilayers. The provided insights present a major step in microscopic understanding of the technologically important charge transfer process in van der Waals heterostructures. 
\end{abstract}
 
\maketitle
 
Transition-metal dichalcogenides (TMDs) have been in the focus of current research due to their enhanced light-matter and Coulomb interaction leading to a rich energy landscape of tightly bound excitons \cite{he2014tightly, chernikov2014exciton, wang2018colloquium,mueller2018exciton}. Stacking TMD monolayers into van der Waals heterostructures introduces spatially separated interlayer states adding another exciton species with long lifetimes and an out-of-plane dipole moment \cite{rivera2015observation,miller2017long,kunstmann2018momentum, jin2019observation,tran2019evidence,seyler2019signatures,alexeev2019resonantly,ruiz2019interlayer, sigl2022optical, holler2022interlayer}. Recent experiments demonstrated the ultrafast charge transfer in optically excited TMD heterobilayers resulting in a formation of interlayer states on a sub-picosecond timescale \cite{hong2014ultrafast,ceballos2014ultrafast, ji2017robust, merkl2019ultrafast, schmitt2021formation, thygesen18}. Typically, TMD heterobilayers exhibit a type-II band alignment \cite{hill2016band, ozcelik2016band} favoring the tunneling of an electron or hole into the opposite layer. However, the underlying microscopic nature of the charge transfer process has not yet been well understood.
In an early previous work, we have suggested a defect-assisted interlayer tunneling directly at the K point \cite{ovesen2019interlayer, merkl2019ultrafast}. Alternatively, a phonon-mediated charge transfer could occur involving intervalley scattering to the strongly hybridized $\Lambda$ or $\Gamma$ valleys \cite{wang2017interlayer,zheng2017phonon, liu2020direct}. A sophisticated microscopic model of such a phonon-assisted formation of interlayer excitons is still missing.

\begin{figure}[t!]
  \centering
  \includegraphics[width=\columnwidth]{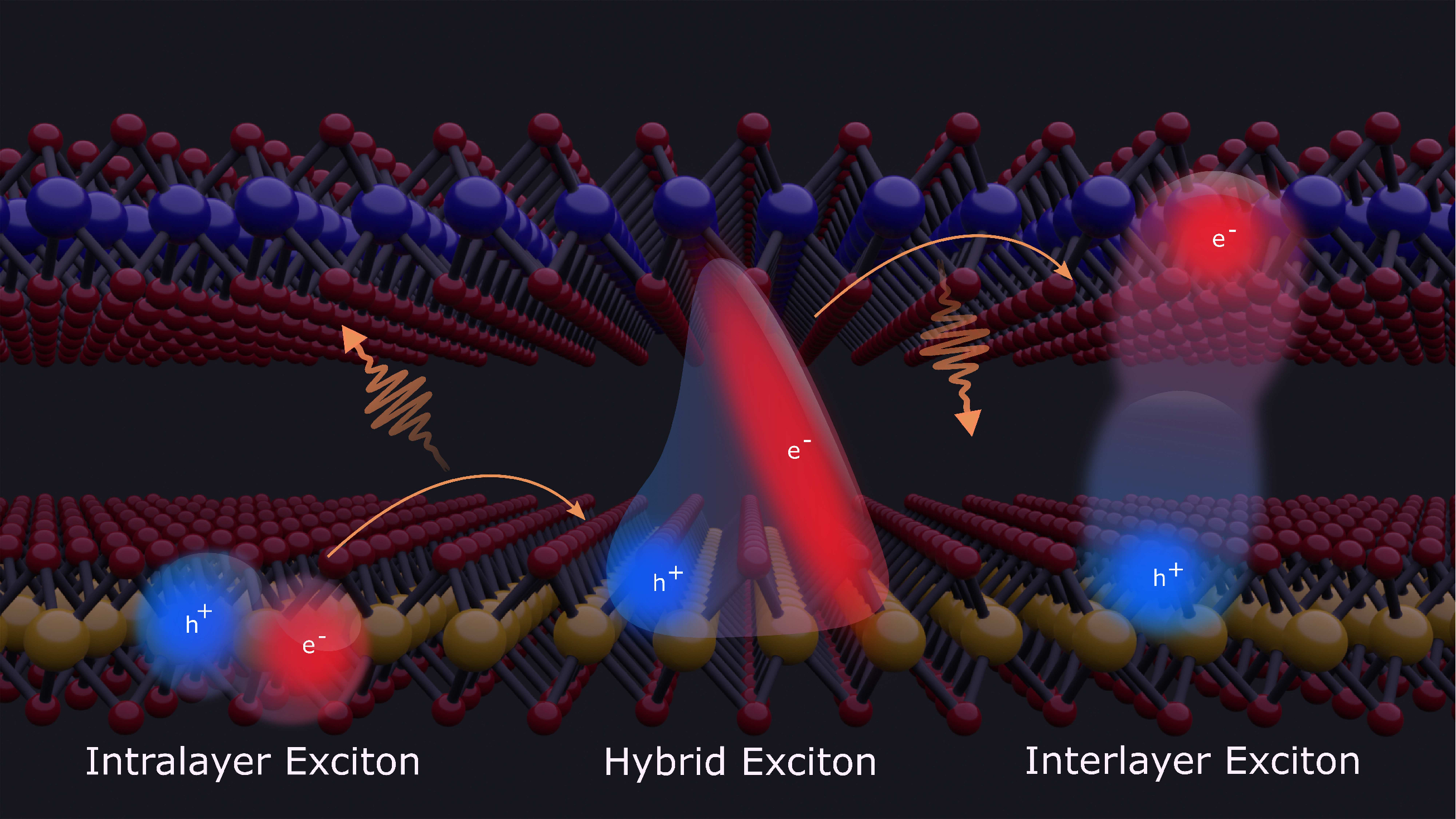}
  \caption{Sketch of the charge transfer process. Starting from an  exciton localized in the bottom layer,  phonon-mediated scattering to an hybrid exciton state (where e.g. the electron lives in both layers) allows for the transfer of the charge (here electron) to the upper layer resulting in a spatially separated interlayer exciton state. In analogy, hole transfer can also take place if hybrid excitons with delocalized holes are present.}
   \label{fig:scheme}
\end{figure}

In this work, we address this open question and reveal the crucial many-particle mechanism behind the ultrafast charge transfer in TMD heterostructures. To this end, we combine first-principle calculations \cite{hagel2021exciton} with the excitonic density matrix formalism \cite{brem2020tunable, katsch2018theory} to obtain a material-realistic model of the excitonic energy landscape, the internal substructure of different exciton species and the phonon-mediated scattering into  layer-hybridized dark intervalley states \cite{brem2020hybridized, merkl2020twist}. We first calculate the exciton energy landscape of the exemplary  MoS$_2$-WS$_2$ and MoSe$_2$-WSe$_2$ heterostructures by solving the Wannier equation for perfectly layer-polarized intra- and interlayer excitons and subsequently computing hybrid excitons based on first-principle interlayer tunneling parameters \cite{ovesen2019interlayer, merkl2019ultrafast, brem2020hybridized, merkl2020twist, brem2020tunable}. Then, we develop and numerically solve equations of motion describing the time- and momentum-resolved evolution of hybrid excitons. This allows us to track the relaxation dynamics of excitons from optically excited intralayer excitons towards charge separated interlayer exciton states. We identify the phonon-mediated intervalley scattering from intralayer KK into the strongly hybridized K$\Lambda^{\prime}$ excitons, followed by the relaxation into energetically lower interlayer KK$^{(\prime)}$ states, as the crucial mechanism behind the ultrafast charge transfer in these heterostructures, cf. Fig \ref{fig:scheme}.  We further determine the characteristic temperature- and stacking-dependent charge transfer time that can  guide future experiments investigating interlayer excitons in van der Waals heterostructures.      

\textit{Microscopic approach:}
The starting point of this work is the Hamilton operator describing electrons and holes of the heterostructure in the basis of monolayer eigenstates (localized in one of both layers). Here we include a stacking-dependent alignment shift of the two monolayer band structures \cite{kormanyos2015k} as well as interlayer tunneling terms resulting from the wave function overlap between the adjacent layers. The necessary material-specific parameters have been extracted from first-principle calculations \cite{hagel2021exciton}. Moreover, we include many-particle interaction Hamiltonians, such as electron-light and electron-phonon coupling as well as the Coulomb interaction between electrons and holes. Here, the scattering between electrons and photons/phonons preferably occurs locally within one of the two layers, whereas we explicitly include the Coulomb interaction between particles residing in different layers. 
The different intra- and interlayer Coulomb matrix elements are computed with a modified Keldysh-type potential \cite{ovesen2019interlayer, brem2020hybridized, brem2020tunable}  accounting for the dielectric environment created by the TMD layers  and the substrate \cite{laturia2018dielectric}.
To achieve a numerically feasible model we set the twist-angle between the two monolayers to zero and study the charge transfer in a spatially homogeneous system characterized by a single atomic alignment. 
Although the twist-angle is known to have a large impact on the hybridization of exciton states \cite{brem2020hybridized}, we expect the qualitative charge transfer behaviour to remain the same also in twisted heterostructures.
Moreover, we do not consider spin-flipping processes and restrict our model to the optically active (A exciton) spin configuration, as the spin-flipping processes are expected to occur on a slower timescale  \cite{song2013transport,glazov2014exciton}.

Now, we derive the dynamics of the system by initially performing a series of basis transformations. First we solve the Wannier equation for pure intra- or interlayer excitons \cite{ovesen2019interlayer} and use the eigenfuctions $ \psi^{\mu}({\bf k})$ to introduce a new set of excitonic operators \cite{brem2020hybridized}  ${X^{\mu\dagger}_{{\bf Q}}} = \sum_{\bf k}{  \psi^{\mu}({\bf k}) a^{\dagger}_{c, \zeta_e, L_e,{\bf k}+\alpha{\bf Q} } a^{}_{v,\zeta_h, L_h,{\bf k}-\beta{\bf Q}}}$ with  the compound quantum number $\mu = (n, \zeta,L)$ labelling the excitonic states. Here $n$ is associated to the series of Rydberg-like states determining the relative electron-hole motion, $\zeta = (\zeta_e,\zeta_h)$ denotes the electron and hole valleys and the layer compound index $L=(L_e,L_h)$ contains the electron and hole layer. Furthermore, we have introduced the center-of-mass momentum $\bf Q$ and the relative momentum $\bf k$ between electrons and holes. The operator $a^{(\dagger)}_{i}$ is annihilating (creating) electrons with the set of quantum numbers denoted by $i$.
We use the new exciton operators to perform a basis transformation to obtain an effective single-particle Hamiltonian for excitons, reading 
\begin{equation}\label{eq:H_x}
    H_\text{X}=\sum_{\mu {\bf Q}} E^{\mu }_{{\bf Q}} X^{\mu\dagger}_{{\bf Q}} X^{\mu}_{{\bf Q}}+ \sum_{\mu\nu{\bf Q}} \Tb_{\mu \nu}  X^{\mu\dagger}_{{\bf Q}}X^{\nu}_{{\bf Q}} 
\end{equation}
with the exciton energy $ E^\mu_{\bf Q}$ obtained from the Wannier equation and the excitonic tunnelling matrix elements $\Tb_{\mu \nu}$, which  contain apart from electronic tunneling rates also the overlap of excitonic wave functions.

\begin{figure}[t!]
  \centering
  \includegraphics[width=\columnwidth]{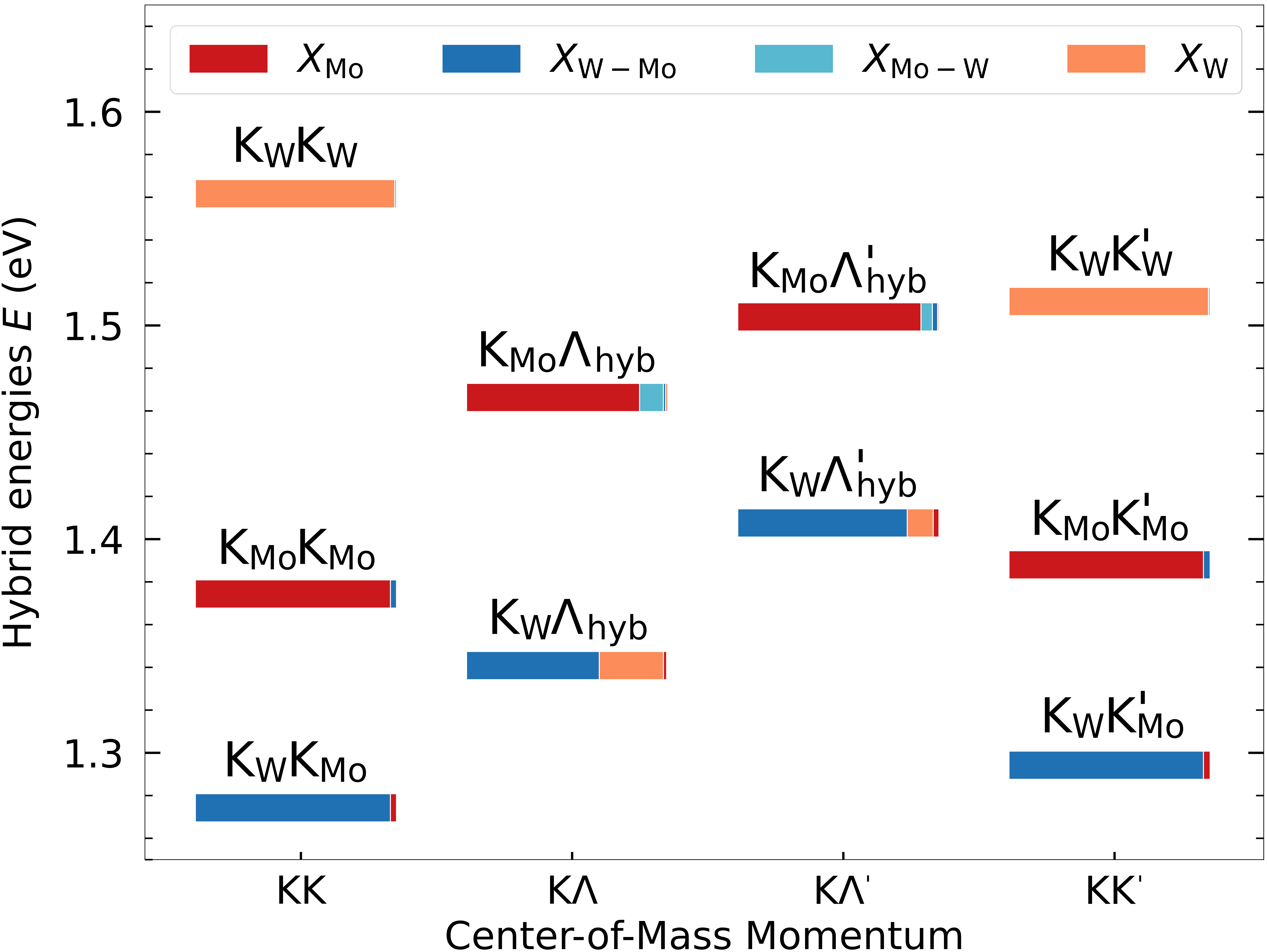}
  \caption{Hybrid-exciton energy landscape for MoSe$_2$-WSe$_2$ ($R^h_h$ stacking).  We use different colors for depicting the four initial intra- and interlayer excitonic states named with $X_{l_h-l_e}$ (using only one index for intralayer excitons). 
  The final hybrid exciton states are denoted with two capital letters (K, $\Lambda$) describing the valley and the subscripts (W, Mo) describing the layer, in which the hole (first letter) and electron (second letter) are localized.
  We highlight for each hybrid exciton the percentage of the involved intra- and interlayer exciton states. Due to the strong tunneling experienced by electrons, the states in the K$\Lambda^{(\prime)}$ valleys are strongly hybridized. Note that we plot only a selection of low-energy hybrid exciton states contributing directly to the relaxation dynamics.}
  \label{fig:landscape}
\end{figure}

Next, we diagonalize the exciton Hamiltonian Eq. \ref{eq:H_x} by introducing a new set of operators 
 $   Y^\eta_{\bf Q} = \sum_{ \mu }{ c^\eta_\mu({\bf Q}) X^\mu_{\bf Q} } 
$ describing hybrid excitons. These are layer-hybridized states consisting of intra- and interlayer excitons with the mixing coefficients $c^\eta_\mu({\bf Q})$ and the  new quantum number $\eta$ defining the hybrid-exciton bands.
The diagonalized Hamiltonian reads in this basis $H_\text{Y} = \sum_{\eta}{\mathcal{E}^\eta_{\bf Q}  Y^{\eta \dagger}_{\bf Q}  Y^\eta_{\bf Q} }$ with the corresponding hybrid-exciton energies $\mathcal{E}^\eta_{\bf Q}$. With the procedure described above we have a microscopic access to the full spectrum of strongly or weakly hybridized exciton states including bright KK as well as momentum-dark intervalley states \cite{selig2018dark, thygesen19}, such as K$\Lambda^{\prime}$ and KK$^{\prime}$, cf. Fig. \ref{fig:landscape} that will be discussed further below.

Finally, we consider the interaction of hybrid exciton states with phonons. As we restrict our study to the low-density regime, exciton-exciton scattering can be neglected \cite{erkensten21}. Starting from the electron-hole picture and performing the same change of basis as described above, we obtain the following Hamiltonian for the hybrid-exciton-phonon interaction \cite{brem2020hybridized}
\begin{equation}\label{eq:1}
   H_\text{Y-ph} = \sum_{\substack{{\bf Q},{\bf q},\\j, \eta,\xi}}{ \Da^{\xi\eta}_{j{\bf q}{\bf Q}} Y^{\xi\dagger }_{{\bf Q + q}}Y^{\eta}_{{\bf Q}} b_{j,{\bf q}} } + h.c.
\end{equation}
 as well as for the hybrid-exciton-light coupling $H_\text{Y-l} = \sum_{\sigma,{\bf Q},\eta}{ {\bf A} \cdot \mathcal{M}^\eta_{\sigma {\bf Q}} Y^\eta_{\bf Q_{\parallel} }} + h.c$. All details on the basis transformation and the resulting hybrid matrix elements can be found in the SI.

Having determined the Hamilton operator $H= H_\text{Y} + H_\text{Y-ph} + H_\text{Y-l}$ for hybrid-excitons and their interaction with phonons and light, we can now derive equations of motion describing the exciton dynamics. Here we exploit the Heisenberg equation of motion  for the occupation numbers $N^\eta_{\bf Q} =  \expval*{Y^{\eta \dagger}_{\bf Q}  Y^\eta_{\bf Q}}$, truncating the Martin-Schwinger hierarchy arising from the exciton phonon-scattering within the second-order Born-Markov approximation \cite{haug2009quantum, thranhardt2000quantum,selig2018dark,brem2018exciton}. 
Considering separately  the coherent polarization $P^{\eta}_{\bf Q} = \expval*{Y^{\eta \dagger}_{\bf Q}}$ and the purely incoherent population $
    \delta N^{\eta }_{\bf Q} = N^{\eta }_{\bf Q} - \abs*{P^{\eta}_{\bf Q}}^2$, we arrive at the semiconductor Bloch-equations in hybrid-exciton basis
\begin{align}\label{eq:15}
    &i\hbar \dot{P}^\eta_0 = -(\mathcal{E}^\eta_0 + i \Gamma^\eta_0)P^\eta_0 -  \mathcal{M}^\eta_0 \cdot {\bf A}(t)\\[5pt]
    &\delta \dot{N}^\eta_{\bf Q} = \sum_{\xi}{ W^{\xi\eta}_{{\bf 0 Q}}  \abs{P^{\eta}_0}^2 } + \sum_{\xi, {\bf Q'}}{ \left( W^{\xi\eta}_{{\bf Q' Q}} \delta N^\xi_{\bf Q'} - W^{\eta\xi}_{{\bf Q Q'}} \delta N^\eta_{\bf Q} \right) } \nonumber.
\end{align}
The details on the scattering tensor $W^{\eta\xi}_{{\bf Q Q'}}$ can be found in the SI.
Equation (\ref{eq:15}) provides full microscopic access to the dynamics of hybrid excitons including optical excitation as well as phonon-scattering-induced relaxation across intra- and intervalley as well as intra- and interlayer states, effectively giving rise to a multi-step charge transfer process.

\section{Hybrid exciton landscape}
We focus here on the two most studied heterostructures in literature, MoS$_2$-WS$_2$ and MoSe$_2$-WSe$_2$. For simplicity, we show the results for the latter in the main text and the former in the SI. 
We start by presenting the hybrid exciton landscape that has been calculated by solving the Wannier equation in the hybrid-exciton basis, cf. Fig. \ref{fig:landscape}. This energy landscape is the key to understanding the charge transfer process. We  use the following notation for the hybrid exciton states: the capital letters describe the valley and the subscript the layer, where the first letter denotes  the hole and the second the electron. To give an example,  K$_\text{W}$K$^\prime_\text{Mo}$ means that the hole is located at the K point of the WSe$_2$ layer, while the electron is localized at the K$^\prime$ valley of the MoSe$_2$ layer.  Furthermore, we use the subscript $\text{hyb}$  to underline that the electron/hole in the corresponding valley is strongly hybridised between both layers, e.g. in  K$_\text{W} \Lambda_\text{hyb}$ the electron at the $\Lambda$ valley lives in both layers.

Figure \ref{fig:landscape} shows the energy landscape of hybrid-excitons in the MoSe$_2$-WSe$_2$ heterostructure for the 
case of R$^h_h$ stacking, i.e. the metal atoms of one layer are placed on top of the metal atoms of the other layer. 
The corresponding landscape for the  other two high-symmetry stackings R$^X_h$ and R$^M_h$ \cite{hagel2021exciton} (where either the chalcogen atom X or the metal atom M of the upper layer is above the hole/void of
the other layer) as well as for the MoS$_2$-WS$_2$ heterostructure can be found in the SI.
We show only the hybrid exciton states that are energetically close to or lower than the intralayer K$_\text{W}$K$_\text{W}$ exciton in the WSe$_2$ layer, since we will resonantly excite the material at this exciton energy and phonon-driven relaxation processes will distribute the excitons towards lower energies. We have checked that the contribution of higher exciton states  to the relaxation dynamics and the charge transfer process, i.e. due to absorption of phonons, is negligible. Note that for this particular heterostructure $\Gamma$K excitons do not play a role for the charge transfer process, while their are crucial for the MoS$_2$-WS$_2$ heterostructure considered in the SI.

In the exciton basis, the hybridization
of electronic states corresponds to a mixing of intra- and interlayer
excitons. We quantify the contribution of each state to the new hybrid-exciton states  by evaluating the mixing coefficients. Here, $|c^\eta_\mu({\bf Q})|^2$ can be interpreted as the percentage of the exciton state $\mu$ inside the hybrid state $\eta$. In the presence of strong tunnelling the new hybrid states are expected to be heavily influenced by different excitonic species. In contrast,  a weak tunnelling should result in hybrid states that are almost purely intra- or interlayer excitons. The degree of hybridisation of each state is illustrated in Fig. \ref{fig:landscape} 
by adopting a color scheme, where we highlight for each hybrid state the different exciton contributions. Here, a hybrid state of a pure intralayer or interlayer character is just red or blue, respectively.  In contrast, strongly hybridised states consist of different colors. Figure \ref{fig:landscape} illustrates that hybrid states involving excitons at the $\Lambda$ valley (K$_\text{W} \Lambda_{\text{hyb}}$, K$_\text{W} \Lambda^\prime_{\text{hyb}}$) contain large contributions of several species, whereas the states at the K valley are either intra- or interlayer excitons to a very high percentage. The weak hybridisation of KK excitons is well known in literature \cite{cappelluti2013tight,gillen2018interlayer}. The
electronic wave functions at the K valley are mostly composed of d orbitals localized at transition metal atoms, which are sandwiched
by the selenium atoms preventing an efficient overlap of wave functions. In contrast, the electronic wave function at the $\Lambda$ valley  has large contributions at
the selenium atoms resulting in much more efficient hybridisation of K$_\text{W} \Lambda^{(\prime)}_{\text{hyb}}$ states \cite{cappelluti2013tight, gillen2018interlayer,merkl2020twist, brem2020hybridized}. 

The energetically lowest states in the investigated MoSe$_2$-WSe$_2$ heterostructure are K$_\text{W}$K$^{(\prime)}_\text{Mo}$ excitons that are almost purely of interlayer exciton character (blue). When exciting the material resonantly to the intralayer K$_\text{W}$K$_\text{W}$ state (orange), there is a number of spectrally lower-lying states that will give rise to a phonon-mediated cascade of transitions down to the energetically lowest states. Note that the scattering process between two hybrid states requires that the initial and final states live at least partially in the same layer. Therefore, we expect the strongly hybridised exciton states K$_\text{W} \Lambda^{(\prime)}_{\text{hyb}}$ to play a major role for the relaxation dynamics and the charge transfer process.

\section{Hybrid exciton dynamics}
Now, we investigate the time- and momentum-resolved relaxation cascade of hybrid excitons after an optical excitation resonant to the purely intralayer K$_\text{W}$K$_\text{W}$ exciton localized in the WSe$_2$ layer, cf. Fig. \ref{fig:landscape}. To focus on the charge transfer process and to avoid interplay effects with the exciting laser pulse, we directly initialize the system with a population in the K$_\text{W}$K$_\text{W}$ state. We have also performed calculations including the laser pulse and the interference of optical excitation and relaxation dynamics, which are presented in the SI. 
Note that we focus on the 1s ground state for all exciton species, as higher-energy states in the Rydberg-like series of excitons exhibit a much smaller scattering probability compared to the 1s-1s transitions \cite{brem2019intrinsic}. This has been verified by numerically evaluating phonon-assisted scattering involving higher-energy states.

\begin{figure}[t!]
  \centering
  \includegraphics[width=\columnwidth]{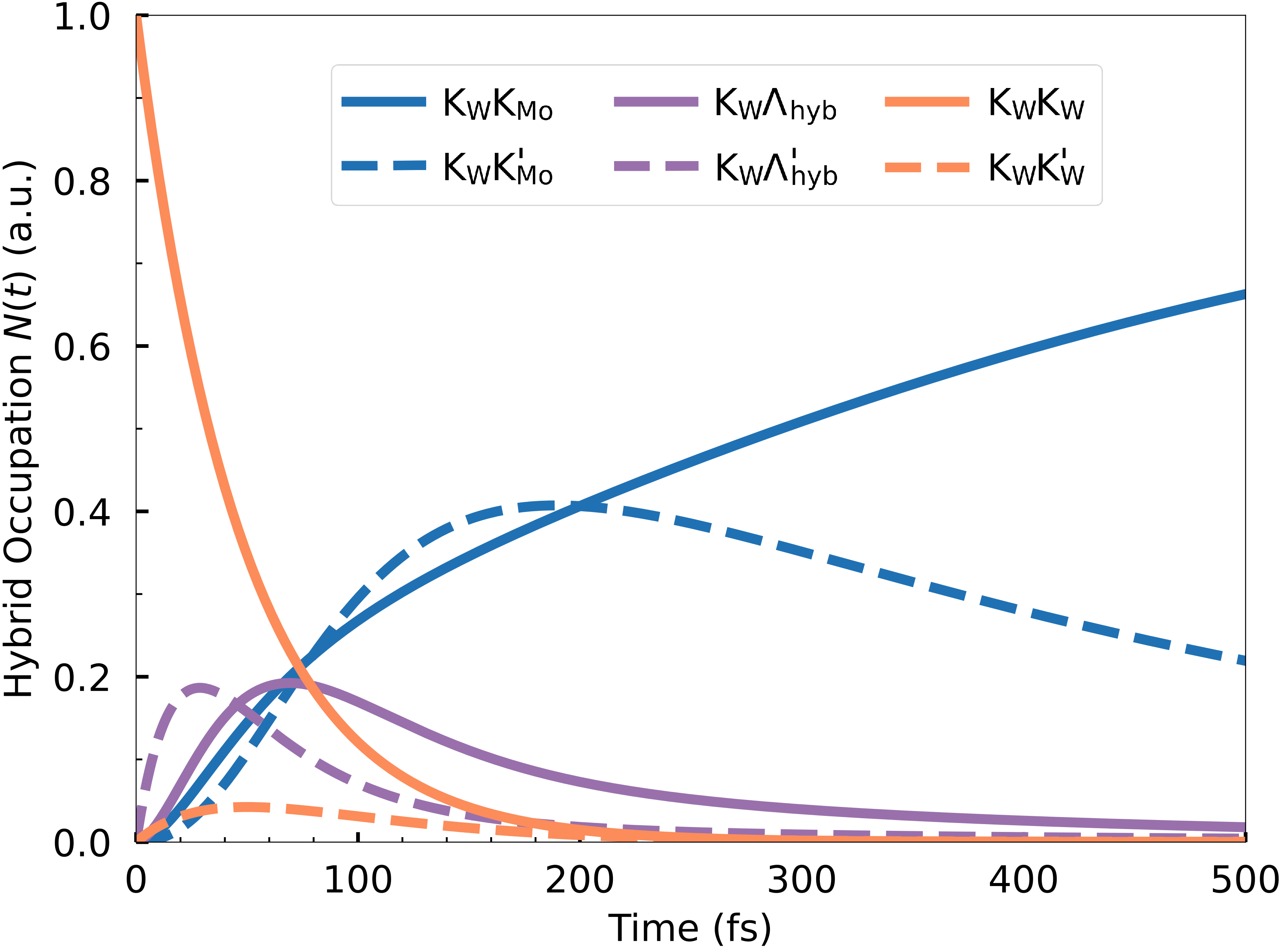}
  \caption{Momentum-integrated hybrid-exciton dynamics at 77 K for MoSe$_2$-WSe$_2$ in R$^h_h$ stacking. By solving Eq. (\ref{eq:15}), we have microscopic access to the  phonon-mediated relaxation dynamics of hybrid exciton and the resulting charge transfer mechanism. Starting with an initial occupation of intralayer K$_\text{W}$K$_\text{W}$ excitons localized in the WSe$_2$ layer (orange line) via phonon-mediated scattering into the strongly hybridized K$_\text{W}$ $\Lambda^{(\prime)}_{\text{hyb}}$ states (purple lines), we end up in the energetically lowest interlayer K$_\text{W}$ K$^{(\prime)}_\text{Mo}$ excitons (blue lines), i.e. the electron has been transferred to the  MoSe$_2$ layer.  }
  \label{fig:dynamics_snapshtos}
\end{figure}

Evaluating the semiconductor Bloch equations (cf. Eq. (\ref{eq:15}) ), we have full microscopic access to the time-, energy- and momentum-resolved relaxation cascade of non-equilibrium excitons. Figure \ref{fig:dynamics_snapshtos} shows the momentum-integrated exciton dynamics in MoSe$_2$-WSe$_2$ (in R$^h_h$ stacking)  at 77 K. We see a decrease of the initially populated intralayer K$_\text{W}$K$_\text{W}$ exciton state (solid orange line). At the same time, we find an ultrafast increase in the population of the hybrid
 K$_\text{W}\Lambda^{(\prime)}_{\text{hyb}}$ excitons on a timescale of sub-100fs (solid and dashed purple lines).
  The microscopic origin of this efficient scattering lies in the nature of the hybrid-exciton-phonon coupling. Phonons can only couple states that share the same layer quantum number $L = (L_e,L_h)$ as exciton-phonon scattering is considered to be a local process. For this reason, phonons can couple pure intra- and interlayer states only through scattering via hybrid states. Once the electron/hole has been scattered into a hybridized state, i.e. into a superposition between both layers, there is a non-zero probability of further scattering into the opposite layer. 
  
Following the relaxation cascade, we can track the population transfer from the hybridised K$_\text{W}\Lambda^{(\prime)}_{\text{hyb}}$ to the interlayer K$_\text{W}$K$^{(\prime)}_\text{Mo}$ excitons (solid and dashed blue lines in Fig. \ref{fig:dynamics_snapshtos}).
After 100 fs, the initially populated 
intralayer K$_\text{W}$K$_\text{W}$ exciton states has been almost completely emptied and most occupation is found in the interlayer K$_\text{W}$K$^{(\prime)}_\text{Mo}$ excitons, where electrons and holes are  spatially separated. As a result, the transfer of electrons from the initial WSe$_2$ layer into the opposite MoSe$_2$ layer occurs on sub-100fs timescale.

\begin{figure}[t!]
  \centering
  \includegraphics[width=\columnwidth]{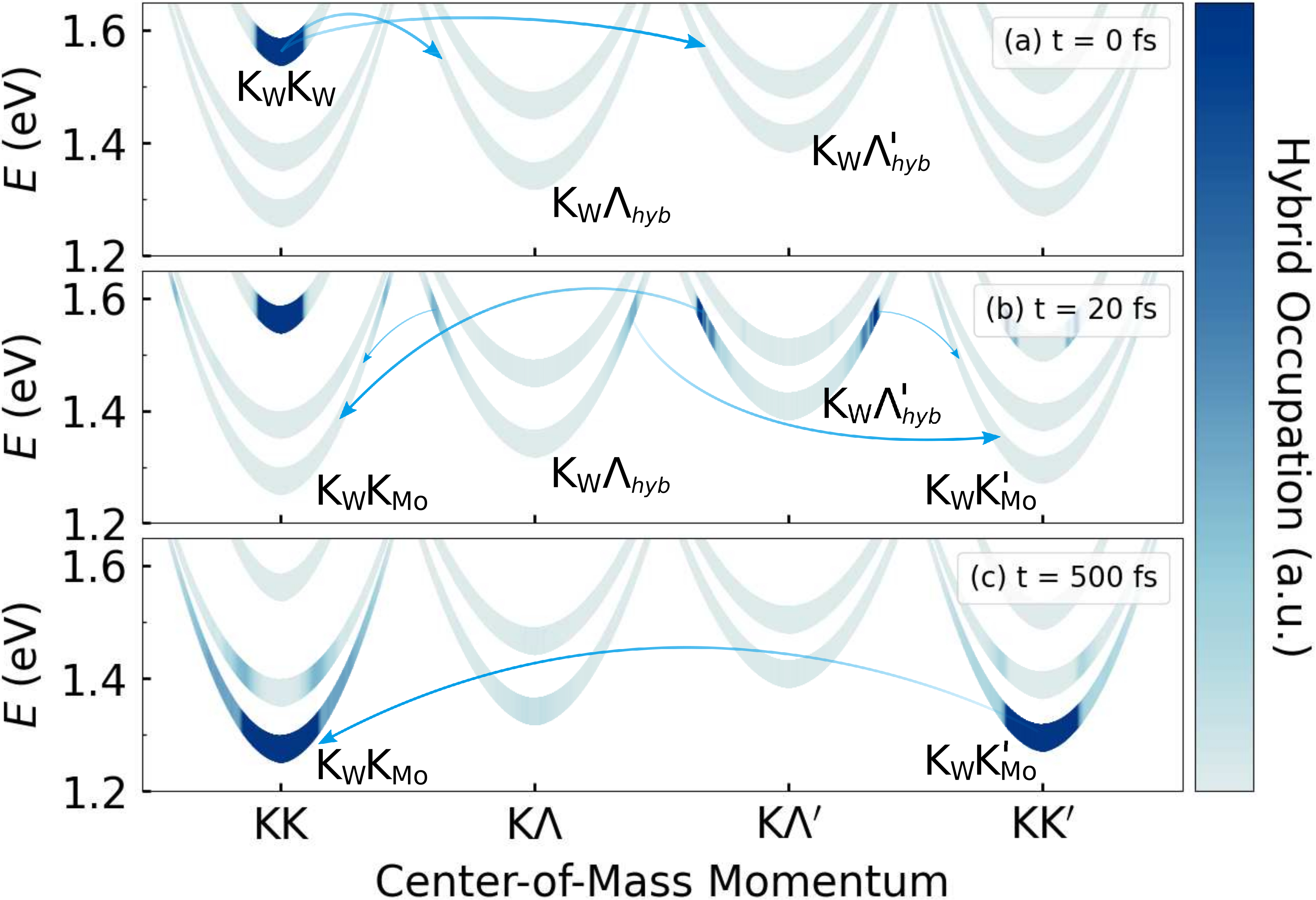}
  \caption{Momentum-resolved hybrid-exciton dynamics  at  (a) 0 fs, (b) 20 fs, and (c) 500 fs. Starting from a population created in the intralayer K$_\text{W}$K$_\text{W}$ exciton,  we highlight the most important phonon-driven scattering processes.  Note that the blue-shading in the parabolas corresponds to a microscopically calculated exciton occupation. 
   The charge transfer of electrons occurs in a two-step process with an initial partial transfer into the hybrid K$_\text{W}$ $\Lambda^{(\prime)}_{\text{hyb}}$ exciton states (with the electron living in both layers) followed by the complete transfer to the energetically lower interlayer K$_\text{W}$ K$_\text{Mo}^{(\prime)}$ states (with the electron localized in the second layer).}
  \label{fig:dynamics_momentum}
\end{figure}

To further illustrate the main scattering processes governing the relaxation cascade, Fig. \ref{fig:dynamics_momentum} shows the momentum-resolved exciton dynamics for different times. We find that in the first step the hybridised K$_\text{W}\Lambda_{\text{hyb}}$ and K$_\text{W}\Lambda^{\prime}_{\text{hyb}}$ states are not populated (0 fs). The scattering into the latter happens on a faster timescale, as here M phonons are involved which are known to give rise to a very efficient scattering with excitons \cite{jin14}. With some delay, there is an efficient phonon-mediated scattering from these hybridized states into the interlayer  K$_\text{W}$K$_\text{Mo}$ and K$_\text{W}$K$^{\prime}_\text{Mo}$ excitons. The population of the latter occurs faster again due to the involved M phonons. In the final step, this state becomes partially depopulated in favor of the energetically lowest  K$_\text{W}$K$_\text{Mo}$ state. After approximately 500 fs a thermalized exciton distribution is reached with the highest occupation in  K$_\text{W}$K$_\text{Mo}$ followed by a certain thermal occupation in  K$_\text{W}$K$^\prime_\text{Mo}$. All other states have only a negligible population.

So far, we have investigated the simplified situation of an initially populated K$_\text{W}$K$_\text{W}$. In a real experiment, this state will be continuously optically excited throughout a finite time window and there will be an interplay of excitation and phonon-mediated scattering. Evaluating Eq. (\ref{eq:15}) we can resolve this interplay and find the same general behavior as described above, cf. the SI. We  observe the same main relaxation steps and a very similar timescale for the charge transfer mechanism. However, tracking the dynamics becomes more complicated during the initial phase of the relaxation due to the simultaneous pumping of excitons in the system that immediately start to relax very rapidly.
The main difference between the simulation with a pump pulse (Fig. 3 SI) and the instantaneous initialization is that at the time when the laser pulse reaches its maximum  a large fraction of excitons has already relaxed to lower energy states, which quantitatively modifies the delay between peak populations of hybrid and interlayer exciton states.  This suggest that we can capture the main features of the process using instantaneous excitation which allows us to gain a much more intuitive picture of the charge transfer without losing generality.

\section{Interlayer charge transfer}

\begin{figure}[t!]
  \centering
  \includegraphics[width=\columnwidth]{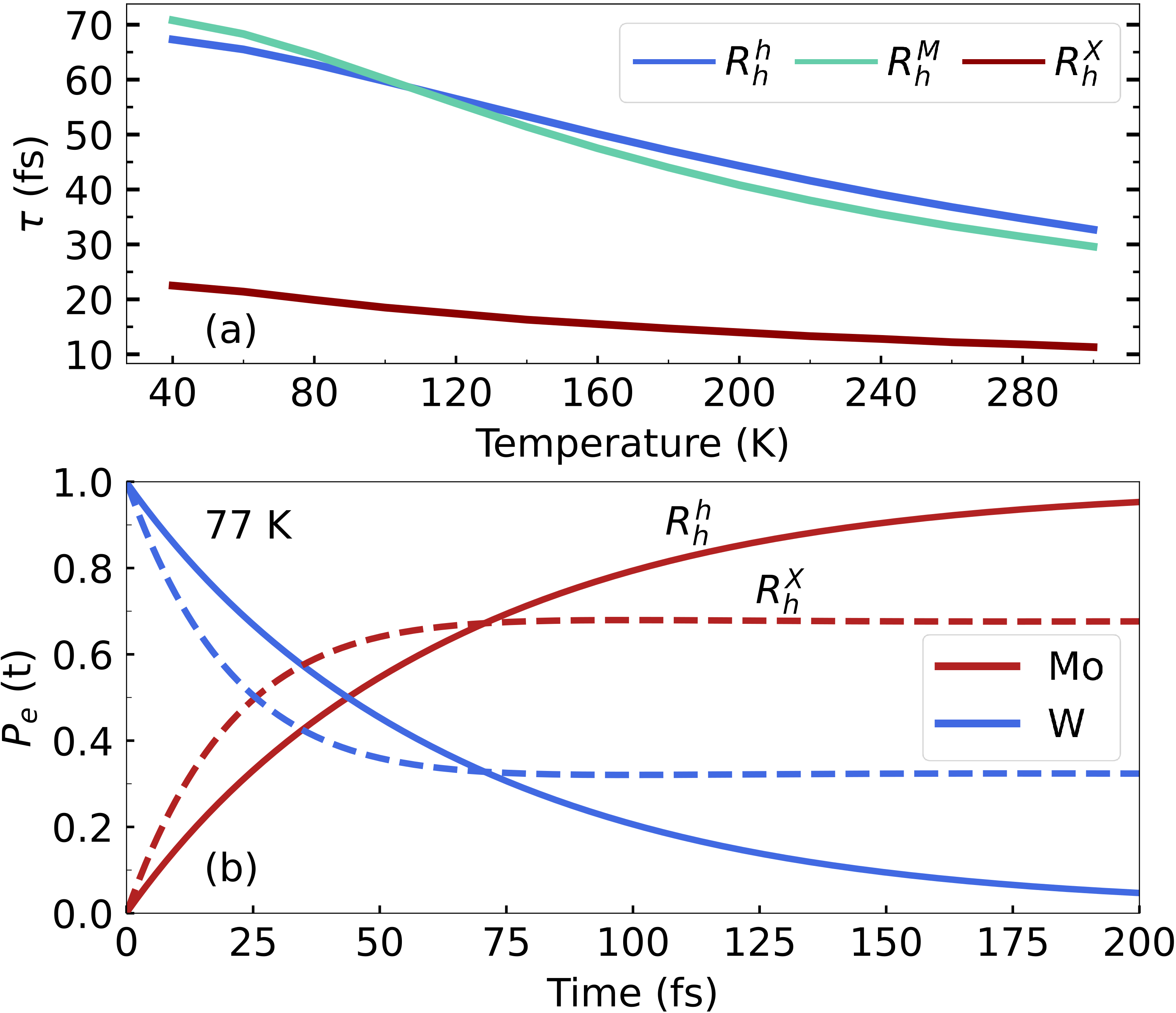}
  \caption{(a) Characteristic electron transfer time as a function of temperature for MoSe$_2$-WSe$_2$ in  different high-symmetry stackings. The time is extracted from an exponential fit of the layer-dependent electron probability $P_e (t)$  as shown in part (b). We find a considerable  decrease in the charge transfer time with  temperature reflecting a more efficient exciton-phonon scattering. Interestingly, we predict a much faster transfer time for R$^X_h$ stacking, as here the hybrid $\Lambda^{(\prime)}_{\text{hyb}}$ states are very close to the interlayer K$_\text{W}$K$_\text{Mo}^{(\prime)}$ states, cf the sI. The faster electron transfer speed comes at the cost of a more incomplete transfer process as the stationary occupation of the $\Lambda^{(\prime)}_{\text{hyb}}$ excitons is relatively high, where the electrons is delocalized between both layers, cf. the dashed vs solid lines in part b. }
  \label{fig:charge_transfer_time}
\end{figure}

Summarizing the exciton dynamics in a nutshell, the initially inserted occupation of the intralayer K$_\text{W}$K$_\text{W}$ excitons is distributed to the energetically lowest interlayer  K$_\text{W}$K$^{(\prime)}_\text{Mo}$ states through an intermediate step involving strongly hybridized K$\Lambda/\Lambda^{(\prime)}$ states. 
This means that the charge transfer is a two-step process, where the electron is first transferred into a hybrid state (representing a superposition of both layers) and in a second step it is transferred to the opposite layer. The characteristic charge transfer time $\tau$ is illustrated in Fig. \ref{fig:charge_transfer_time}(a) as a function of temperature for different high-symmetry stackings.
We can quantify the charge transfer speed by computing the layer- and stacking-dependent probability $P_e(t)=\langle a^\dagger_ca^{}_c\rangle$ of one electron  being localized in the MoSe$_2$ layer after excitation of an intralayer state in the WSe$_2$ layer, cf. Fig. \ref{fig:charge_transfer_time}(b). 
By exponentially fitting the temporal evolution of $P_e(t)$, we can extract the characteristic electron transfer time $\tau$.  We find an ultrafast transfer rate of $\tau=33$ fs  for MoSe$_2$-WSe$_2$ in R$^h_h$ stacking at room temperature. The electrons are almost completely transferred from the initially occupied WSe$_2$ layer to the MoSe$_2$ layer, i.e. one finds the electron with a probability of 95\% after 200 fs, cf. the solid red and blue lines in Fig. \ref{fig:charge_transfer_time}(b).
 
Since the relaxation cascade is mediated by phonons, we find a pronounced temperature dependence of the transfer time. Concretely, we predict an increase in $\tau(T)$ by approximately a factor of 2 to $\tau=67$ fs at 40 K for R$^h_h$ stacking. The reason is the reduced scattering efficiency with phonons at lower temperatures. Nevertheless, even at cryogenic temperatures we find an ultrafast charge transfer as the relaxation cascade occurs toward energetically lower exciton states and is driven by phonon emission. 

Interestingly, we find an unexpected acceleration of the charge transfer for R$^X_h$ stacking (whereas R$^M_h$ stacking is rather similar to the R$^h_h$ stacking investigated so far).
This originates from the hybrid-energy landscape for different stackings (cf. the SI). 
The stronger tunnelling at the K$\Lambda$ valley for the R$^X_h$ stacking \cite{hagel2021exciton} and the resulting  larger red-shift of exciton energies has as a consequence that the relevant energy levels are closer than in other stackings, cf. Fig. 2(a) in the SI. In particular,  the strongly hybridised K$\Lambda_{\text{hyb}}$ states and the  interlayer K$_\text{W}$K$_\text{Mo}^{(\prime)}$ excitons are nearly degenerate. As a result, the second step in the charge transfer process is much more efficient compared to the R$^h_h$ stacking.
Note however that while the charge transfer is indeed faster for  the $R^X_h$ stacking, there is only an incomplete transfer. This means that the electron is not transferred to almost 100\% as in the case of R$^h_h$ stacking, but there is still a probability of approximately 40\% to find the electron in the initially populated layer, cf. the dashed lines in Fig. \ref{fig:charge_transfer_time}(b). The reason behind this is that a large percentage of the hybrid-exciton population remains in the  hybrid K$\Lambda_{\text{hyb}}$ state as it is threefold degenerate and very close in energy with the lowest interlayer K$_\text{W}$K$_\text{Mo}$ state. Hence, the electron remains partially delocalized between the two layers and the charge transfer is incomplete.  

So far we have investigated the MoSe$_2$-WSe$_2$ heterostructure. The comparison with MoS$_2$-WS$_2$ (shown in the SI) yields the same general behavior for the hybrid-exciton relaxation dynamics. We find a somewhat slower charge transfer with $\tau = 88$ fs for R$^h_h$ stacking at room temperature, mainly due to the much larger energy window involved in the relaxation dynamics, cf. the energy landscape in Fig. S4 in the SI.  Analyzing the results in more detail,  we find the main difference originating from the importance of $\Gamma_{\text{hyb}}$K excitons. The strong tunnelling occurring in the $\Gamma$ valley results in a large red-shift of the corresponding exciton states making them energetically lowest in MoS$_2$-WS$_2$. 
Interestingly, we find that in contrast to  MoSe$_2$-WSe$_2$ discussed above,  we find here the slowest charge transfer for the R$^X_h$ stacking. This can be traced back to $\Gamma_{\text{hyb}}$K$_\text{W}$ 
states which trap excitons as the only states that are close in energy and have a similar composition are K$\Lambda^{(\prime)}_{\text{hyb}}$ excitons. However,  scattering into the latter from $\Gamma_{\text{hyb}}$K$_\text{W}$  requires a simultaneous electron and hole transfer and is thus a negligible higher-order process. A more detailed description of the relaxation dynamics as well as temperature- and stacking-dependent charge transfer times in the MoS$_2$-WS$_2$ heterostructure can be found in the SI.

In conclusion, we have developed a microscopic and material-specific theory allowing us to access the relaxation dynamics of hybrid excitons in van der Waals heterostructures. In particular, we identify the extremely efficient phonon-mediated relaxation via strongly hybridized K$\Lambda_{\text{hyb}}$ excitons as the crucial mechanism behind the ultrafast charge transfer process in the MoSe$_2$-WSe$_2$ heterostructure. We predict charge transfer times in the range of tens of femtoseconds that are strongly dependent on temperature and stacking of the layers. Our work presents an important step towards a microscopic understanding of the relaxation cascade and ultrafast charge transfer in technologically promising van der Waals heterostructures.  \\

We acknowledge support from Deutsche Forschungsgemeinschaft
(DFG) via SFB 1083 (Project B9) and the European
Unions Horizon 2020 Research and Innovation Program, under
Grant Agreement No. 881603 (Graphene Flagship).

\bigskip

{\bf DATA AVAILABILITY STATEMENT}\\
The data that support the findings of this study are available from the corresponding author upon reasonable request.

{\bf CONFLICT OF INTEREST}\\
The authors declare no conflict of interest

\bibliography{references}

\end{document}


\title{Ultrafast phonon-driven charge transfer in van der Waals heterostructures\\
\Large Supplementary Information}

\author{Giuseppe Meneghini}
\email{giuseppe.meneghini@physik.uni-marburg.de}
\affiliation{%
 Department of Physics, Philipps University of Marburg, 35037 Marburg, Germany}%

\author{Samuel Brem}

\affiliation{%
 Department of Physics, Philipps University of Marburg, 35037 Marburg, Germany}%

\author{Ermin Malic}
\affiliation{%
 Department of Physics, Philipps University of Marburg, 35037 Marburg, Germany}
 \affiliation{%
 Department of Physics, Chalmers University of Technology, 41258 Göteborg, Sweden
}%


\maketitle

\section*{Theoretical Approach}
\textbf{Keldysh potential:}
To describe the TMD bilayer system we need to include the correct dielectric screening originating from the presence of two different TMDs monolayers. Choosing a reference at $z=0$ at the interface of the two layers, we can address the position of a charge placed in the middle of one of the two layers (denoted by the index $L =0,1$), i.e. $z= \pm d_L/2$ with the layer width $d_L$, addressing the background dielectric constant (with the subscript $bg$). We can write the general dielectric constant as
\begin{equation}\label{eq:2}
    \epsilon^L (\bf r) = 
    \begin{cases}
    \epsilon^L_{bg} ,& \text{if  } z < -d_0\\
    \epsilon^L_{0}  ,& \text{if  } -d_0 < z < 0\\
    \epsilon^L_{1}  ,& \text{if  }  0 < z < d_1\\
    \epsilon^L_{bg} ,& \text{if  } z > d_1
    \end{cases}
\end{equation}
and solve the Poisson equation for the system using as boundary conditions Eq. (\ref{eq:2}) \cite{ovesen2019interlayer,brem2020tunable}. This way we  obtain an analytical expression for the screened Coulomb matrix element
\begin{align}\label{eq:3}
\begin{split}
     W^{L L'}_{{\bf q}} &= \frac{e_0^2}{2 \epsilon^{}_0 A q \,\epsilon^{}_{L L'}(q)} \\[6pt]
     \epsilon^{}_{L L'}(q) &= 
     \begin{cases}
         \epsilon^{L}_{intra}(q) ,& \text{if  } L = L'\\
         \epsilon^{}_{inter}(q) ,& \text{if  } L \neq L'
     \end{cases}
      \end{split}
\end{align}    
with $\epsilon^{}_{inter}(q) = \kappa_{bg} g^0_q g^1_q f_q \quad\text{and}\quad  \epsilon^{L}_{intra}(q) = \frac{\kappa_{bg} g^{1-L}_q f_q }{ch\left( \delta_{1-L} q /2 \right) h^L_q}$, $\kappa = \sqrt{\epsilon^{\parallel} \epsilon^{\perp}}$, $\delta_L = \alpha_L d_L $, $\alpha = \sqrt{\epsilon^{\parallel}/ \epsilon^{\perp}}$. Here, have introduced the following abbreviations:
\begin{align}   
\begin{split}
    f_q  &= 1 + \frac{1}{2} \left[  \left( \frac{\kappa_{0}}{\kappa_{bg}} + \frac{\kappa_{bg}}{\kappa_{0}}\right) th\left( \delta_0 q \right) + \left( \frac{\kappa_{1}}{\kappa_{bg}} + \frac{\kappa_{bg}}{\kappa_{1}}\right) th\left(\delta_1 q \right)  + \left( \frac{\kappa_{0}}{\kappa_{1}} + \frac{\kappa_{1}}{\kappa_{0}} \right) th\left(\delta_0 q \right) th\left(\delta_1 q \right)  \right] \\[6pt]
    h^L_q &= 1+ \frac{\kappa_{bg}}{\kappa_{L}} th\left(\delta_L q \right) + \frac{\kappa_{bg}}{\kappa_{1-L}} th\left(\delta_{1-L} q /2 \right) + \frac{\kappa_{L}}{\kappa_{1-L}} th\left(\delta_L q \right) th\left(\delta_{1-L} q/2 \right)\\[6pt]
    g^L_q &=  \frac{ch \left( \delta_L q \right)}{ch \left( \delta_{1-L} q /2 \right) \left[ 1 + \frac{\kappa_{bg}}{\kappa_L} th\left( \delta_L q \right) \right]}\\[12pt]
\end{split}
\end{align}

\textbf{Tunnelling Hamiltonian:}
In this section, we provide details on the transformation of the tunneling Hamiltonian into the exciton basis \cite{brem2020hybridized}. Starting from the electron-hole picture we can write the tunnelling Hamiltonian in the following way:
\begin{equation}\label{eq:1}
    T = \sum_{i,j,{\bf k},\lambda}{ T^\lambda_{i j}({\bf k q}) a^\dagger_{\lambda i {\bf k}+{\bf q}} a^{}_{\lambda j {\bf k}}} 
\end{equation}
with $T^\lambda_{i j}({\bf k q}) = (1-\delta_{L_i L_j}) \bra{\lambda i {\bf k}+ {\bf q} } V_0+ V_1 \ket{\lambda j {\bf k} }$, where $\lambda=c, v$ is the band index, $i/j = (L,\zeta)$ and $V_{L}$ (with $L=0, 1$) the electrostatic potentials generated by the two layers. Assuming tight-binding wave functions,  the overlap of electronic wavefunctions is becoming rapidly very small for $q > 0$, and thus justifying the restriction to processes of vanishing momentum transfer $q$. In our effective model, where we describe electrons and holes in proximity of high-symmetry points of the Brillouin zone using an effective mass approximation, this allows only intravalley tunneling. 
We now perform the change of basis into the excitonic picture, as explained in the theory section of the main part yielding: 
\begin{equation}
    T_X = \sum_{\substack{\mu,\nu,{\bf Q}}}{ \Tb_{\mu \nu}  {X^{\mu\dagger }_{{\bf Q}}}X^{\nu}_{{\bf Q}} }.
\end{equation}
We have introduced the excitonic tunnelling matrix elements
\begin{align}
\begin{split}
   \Tb_{\mu \nu} = \delta_{L^{\mu}_h L^{\nu}_h} (1- \delta_{L^{\mu}_e L^{\nu}_e})  &\Tb^c_{\substack{\mu \nu}} - \delta_{L^{\mu}_e L^{\nu}_e} (1- \delta_{L^{\mu}_h L^{\nu}_h})  \Tb^v_{\substack{\mu \nu}}\\[6pt]
    \Tb^c_{\substack{\mu \nu}} = \delta_{\zeta^\mu \zeta^\nu} T^c_{ij} \Fa_{\mu \nu} \quad&\text{and}\quad
    \Tb^v_{\substack{\mu \nu}}  = \delta_{\zeta^\mu \zeta^\nu} T^v_{ij} \Fa_{\mu \nu}
\end{split}
\end{align}
with $\Fa_{\mu \nu} = \sum_{\bf k}{ \psi^{\mu*}_{}({\bf k} )\psi^{\nu}_{}({\bf k} )}$ and $T^{\lambda}_{ij}$ defined as in Eq. (\ref{eq:1}) with $i, j$ describing the electron/hole quantum numbers.\\[12pt]

\textbf{Hybrid-exciton-phonon and hybrid-exciton-light Hamiltonian:}
The contribution of the electron-phonon interaction to the Hamiltonian reads in  second quantization
\begin{align}
\begin{split}
    H_{e-ph} &= \sum_{m,n,\lambda,j}{ D^{\zeta^\lambda_m\zeta^\lambda_n\lambda}_{j,{\bf k}_m-{\bf k}_n} a^{\dagger}_{\lambda,m} a_{\lambda,n} \left(b^{}_{j,{\bf k}_n-{\bf k}_m} + b^\dagger_{j,{\bf k}_m-{\bf k}_n} \right)  }\\[6pt]
    \text{with}\quad  D^{\zeta^\lambda_m\zeta^\lambda_n\lambda}_{j,{\bf q}} &\approx \sqrt{\frac{\hbar}{2 \rho_{L^{ph}_j} A \Omega_{j{\bf q}} } } \Tilde{D} ^{\zeta^\lambda_m\zeta^\lambda_n\lambda}_{j,{\bf q}}\\[6pt]
    \text{and} \quad \Tilde{D} ^{\zeta^\lambda_m\zeta^\lambda_n\lambda}_{j,{\bf q}} &= 
    \begin{cases}
        \Tilde{D}^{\lambda}_\zeta {\bf q} \quad\text{if } \zeta^{\lambda}_m=\zeta^{\lambda}_n = \zeta \quad\text{and } \kappa_j =TA,LA\\[6pt]
        \Tilde{D}^{\lambda}_{\zeta^{\lambda}_m \zeta^{\lambda}_n} \quad\text{else }
    \end{cases}\\
    \text{and}\quad \Omega_{j{\bf q}} &= 
    \begin{cases}
        v_j {\bf q}  \quad\text{if } \kappa_j =TA,LA\\
        \Omega_{_j} \quad\text{else }
    \end{cases}
\end{split}
\end{align}
where $a^{\lambda (\dagger)}_m$ are electron (creation) annihilation operators with $\lambda$ as the band index and $m = ({\bf k}_m, \zeta_m,L_m)$ labelling the different electronic quantum states. Here, ${\bf k}_m$ denotes the electron momentum with respect to the main high-symmetry point labeled by $\zeta_m$. Furthermore, we have introduced  $b^{(\dagger)}_{j,{\bf q}}$ as phonon (creation) annihilation operators with $j = (\kappa_j,\zeta^{ph}_j,L^{ph}_j)$ as compound index including the phonon branch $\kappa_j$, the phonon valley index $\zeta^{ph}_j$ and the phonon layer index $L^{ph}_j$. Finally, $\rho$ denotes the surface mass density of the TMD layer.
The strength of the electron-phonon coupling $\Tilde{D}^{\lambda}_{\zeta^{\lambda}_m \zeta^{\lambda}_n}$, the sound velocities ($v_\alpha $) and energy of optical phonons ($\Omega^\lambda_{\zeta^{ph}_j}$) are taken from DFT calculations \cite{jin14}.

Now, we change into the excitonic basis yielding the exciton-phonon Hamiltonian
\begin{align}\label{eq:12}
\begin{split}
    H_{X-ph} &= \sum_{j,{\bf Q},{\bf q}, \mu,\nu}{ \tilde{\Da}^{\nu\mu}_{j,{\bf q},{\bf Q}} X^{\nu\dagger }_{{\bf Q + q}}X^{\mu}_{{\bf Q}} b_{j,{\bf q}} } + h.c. \\[6pt]
     \text{with} \quad \tilde{\Da}^{\nu\mu}_{j,{\bf q},{\bf Q}} &= D^{\zeta^e_\mu \zeta^e_\nu c}_{j,{\bf q}} \delta_{\zeta^h_\mu \zeta^h_\nu} \delta_{\zeta^e_\nu-\zeta^e_\mu, \zeta^{ph}_j} \delta_{L^e_\nu,L^{ph}_j} \delta_{L^e_\nu,L^e_\mu} \Fa^{\mu \nu}\left(  \beta_\nu \left[ {\bf q} + s_{\mu\nu}{\bf Q}  \right] \right)+\\[6pt]
    &- D^{\zeta^h_\mu \zeta^h_\nu v}_{j,{\bf q}} \delta_{\zeta^e_\mu \zeta^e_\nu} \delta_{\zeta^h_\nu-\zeta^h_\mu, \zeta^{ph}_j} \delta_{L^h_\nu,L^{ph}_j} \delta_{L^h_\nu,L^h_\mu} \Fa^{\mu \nu}\left(  -\alpha_\nu \left[ {\bf q} + s_{\mu\nu}{\bf Q}  \right] \right)\\[6pt]
    \text{and} \quad \Fa^{\mu \nu}\left({\bf q}\right) &=  \sum_{\bf k}{ \psi^{\mu*}({\bf k}+{\bf q} ) \psi^{\nu}({\bf k} )}
\end{split}
\end{align}
with the excitonic eigenfunction $ \psi^{\mu}({\bf k})$   and with  $s_{\mu\nu} = 1-M_\nu/M_\mu$. Since $s_{\mu\nu}{\bf Q}$ is a small vector in comparison to the  phonon momentum in the intervalley scattering process, we can neglect the ${\bf Q}$-dependence  in the form factors $\Fa$.  This implies $\Da^{\nu\mu}_{j,{\bf q},{\bf Q}} \approx \Da^{\nu\mu}_{j,{\bf q}}$.
Finally, we perform the change into the hybrid-exciton basis \cite{brem2020hybridized}, as  introduced in the main text, and we arrive at the following final hybrid-exciton-phonon Hamiltonian
\begin{align}\label{eq:12}
\begin{split}
    &H_{Y-ph} = \sum_{j,{\bf Q},{\bf q}, \eta,\xi}{ \Da^{\xi\eta}_{j,{\bf q},{\bf Q}} Y^{\xi\dagger }_{{\bf Q + q}}Y^{\eta}_{{\bf Q}} b_{j,{\bf q}} } + h.c. \\[6pt]
    &\text{with}\quad \Da^{\xi\eta}_{j,{\bf q},{\bf Q}} = \sum_{\mu,\nu}{ c^{\eta*}_\mu({\bf Q})c^{\xi}_\nu({\bf Q}+{\bf q} ) \tilde{\Da}^{\nu\mu}_{j,{\bf q}}}.
\end{split}
\end{align}

To include an optical excitation with a laser pulse, we use the interband part of the  light-electron interaction Hamiltonian
\begin{equation}
    H_{e-l} = \frac{e_0}{m_0} \sum_{m,n,\sigma} {\bf A}\cdot {\bf M}^{cv}_{mn, \sigma} a^{\dagger}_{c,m} a_{v,n }
\end{equation}
where $e_0,m_0$ are the electron charge and mass, respectively. Furthermore, ${\bf M}^{cv}_{mn, \sigma} = -i \hbar \bra{n_v}\nabla \ket{m_c}$ is the optical matrix element containing the  selection rules for the system with $\sigma$ describing the polarization and ${\bf A}$ the vector potential of the light pulse.
Now, we change to the exciton picture obtaining 
\begin{align}
\begin{split}
    H_{X-l} &= \sum_{\sigma,{\bf Q},\mu}{ {\bf A} \cdot {\bf \mathcal{M}}^\mu_{{\bf Q} \sigma} X^\mu_{{\bf Q}_{\parallel} }} + h.c.\\[6pt]
    \text{with }\quad & {\bf \mathcal{M}}^\mu_{{\bf Q} \sigma } = \frac{e_0}{m_0} \delta_{\zeta^\mu_e \zeta^\mu_h,{\bf KK}} {\bf M}^{cv}_{{\bf Q} \sigma} \sum_{\bf k}{ \psi^{\mu*}({\bf k})}.
\end{split}
\end{align}
It is important to notice that inside ${\bf M}^{cv}_{{\bf Q} \sigma}$ the selection rules imply that the electron and hole have to be localized at the K valley, since photons exhibit only a negligible momentum and can only induce direct transitions. In the hybrid-exciton basis, we obtain  
\begin{align}
    \begin{split}
        H_{Y-l} &= \sum_{\sigma,{\bf Q},\eta}{ {\bf A} \cdot \tilde{\mathcal{M}}^\eta_{{\bf Q}\sigma } Y^\eta_{\bf Q_{\parallel} }} + h.c.\\[6pt]
        \text{with }\quad & \tilde{\mathcal{M}}^\eta_{\sigma {\bf Q}} = \sum_{\mu}{ c^{\eta *}_\mu( {\bf Q} ) {\bf \mathcal{M}}^\mu_{{\bf Q} \sigma }}\\[12pt]
    \end{split}
\end{align}

\textbf{Hybrid-exciton dynamics:}
After truncating the hierarchy problem to include only single-particle and two-particle contributions, we obtain  two coupled equations
\begin{align}
\begin{split}\label{eq:13}
    \partial_t N^\eta_{\bf Q} &= \frac{2}{\hbar} \sum_{\xi,{\bf Q},\pm}{ Im\left(  \Da^{\xi\eta}_{j,{\bf q}} \mathcal{C}^{\eta \xi,\pm}_{j{\bf q}{\bf Q}} \right) }\\[6pt]
    i\hbar \partial_t \mathcal{C}^{\eta \xi,\pm}_{j{\bf q}{\bf Q}} &= \left( \mathcal{E}^{\xi}_{{\bf Q}-{\bf q}} - \mathcal{E}^{\eta}_{\bf Q} \mp \hbar\Omega_{j{\bf q}} \right) \mathcal{C}^{\eta \xi,\pm}_{j{\bf q}{\bf Q}} - \Da^{\xi\eta *}_{j,{\bf q}} \left( \beta^\mp_{{\bf q}} N^{\xi}_{{\bf Q}-{\bf q}} - \beta^\mp_{{\bf q}} N^{\eta}_{{\bf Q}} \right)
\end{split}
\end{align}
where we have defined $\mathcal{C}^{\eta \xi,\pm}_{j{\bf q}{\bf Q}} = \expval{ Y^{\eta\dagger }_{{\bf Q}}Y^{\xi}_{{\bf Q - q}} b^{\dagger /()}_{j,\mp,{\bf q}} }$, and $\beta^\pm_{{\bf q}} = 1/2 \mp 1/2 +  \expval{ b^{\dagger}_{\bf q} b^{}_{\bf q} }$. To obtain the second equation we have neglected non-linear terms in the densities ($\propto N^2$). Now, we apply the Markov approximation \cite{selig2018dark,haug2009quantum,thranhardt2000quantum,brem2018exciton} for solving the equation for $\mathcal{C}^{\eta \xi,\pm}_{j{\bf q}{\bf Q}}$ yielding
\begin{equation}
    \mathcal{C}^{\eta \xi,\pm}_{j{\bf q}{\bf Q}} \approx i\pi \Da^{\xi\eta*}_{j,{\bf q}} \left( \beta^\mp_{{\bf q}} N^{\xi}_{{\bf Q}-{\bf q}}- \beta^\pm_{{\bf q}} N^{\eta}_{{\bf Q}} \right) \delta \left( \mathcal{E}^{\xi}_{{\bf Q}-{\bf q}} - \mathcal{E}^\eta_{\bf Q} \mp   \hbar\Omega_{j{\bf q}} \right)
\end{equation}
which inserted in the first equation of Eq. (\ref{eq:13}) results in the Boltzmann scattering equation
\begin{align}
\begin{split}
    &\partial_t N^\eta_{\bf Q} = \sum_{\xi,{\bf Q'}}{\left( W^{\xi\eta}_{{\bf Q' Q}} N^{\xi}_{\bf Q'} - W^{\eta\xi}_{{\bf Q Q'}} N^{\eta}_{\bf Q} \right) }\\
    & W^{\eta\xi}_{{\bf Q Q'}} = \frac{2\pi}{\hbar} \sum_{j,\pm}{ \left| \Da^{\eta\xi}_{j,{{\bf Q'-Q}}} \right|^2 \left( \frac{1}{2} \pm \frac{1}{2} + n^{ph}_{j,{\bf Q'-Q}} \right) \delta \left( \mathcal{E}^{\xi}_{\bf Q'} - \mathcal{E}^\eta_{\bf Q} \mp   \hbar\Omega_{j{\bf Q'-Q}} \right) }
\end{split}
\end{align}
We introduce the coherent hybrid polarization $P^{\eta}_{\bf Q} = \expval{Y^{\eta \dagger}_{\bf Q}}$  and the incoherent hybrid exciton population 
\begin{equation}
    \delta N^{\eta }_{\bf Q} = \expval{ Y^{\eta\dagger }_{{\bf Q}}Y^{\eta}_{{\bf Q}}}-\expval{Y^{\eta \dagger}_{\bf Q}}\expval{Y^{\eta}_{\bf Q}} = N^{\eta }_{\bf Q} - \abs{P^{\eta}_{\bf Q}}^2.
\end{equation}
As next, we derive the equation of motion for the coherent polarization
\begin{equation}
    i\hbar \partial_t P^\eta_0 = -(\mathcal{E}^\eta_0 + i \Gamma^\eta_0)P^\eta_0 -  \tilde{\mathcal{M}}^\eta_0 \cdot {\bf A}(t)
\end{equation}
where $0$ refers to ${\bf Q} = 0$ due to the condition that the laser pulse only creates hybrid excitons around ${\bf Q} = 0$ at the K valley. 
The incoherent dynamics is given by 
\begin{equation}
    \delta \dot{N}^\eta_{\bf Q} = \partial_t \left( N^\eta_{\bf Q} - \abs{P^\eta_{\bf Q}}^2 \right) = \dot{N}^\eta_{\bf Q} - 2 Re\left\{ \dot{P}^\eta_{\bf Q} P^{\eta *}_{\bf Q} \right\} \delta_{{\bf Q} = 0}
\end{equation}
which yields the equation of motion for incoherent hybrid exciton densities
\begin{equation}
 \delta \dot{N}^\eta_{\bf Q} = \sum_{\xi}{ W^{\xi\eta}_{{\bf 0 Q}}  \abs{P^{\eta}_0}^2 } + \sum_{\xi, {\bf Q'}}{ \left( W^{\xi\eta}_{{\bf Q' Q}} \delta N^\xi_{\bf Q'} - W^{\eta\xi}_{{\bf Q Q'}} \delta N^\eta_{\bf Q} \right) }
\end{equation}
We use the last equation to track the dynamics of hybrid excitons in TMD heterostructures including  the polarization to population transfer creating incoherent excitons as well as phonon-assisted exciton relaxation into an equilibrium Boltzmann distribution.

\begin{figure}[t!]
\begin{subfigure}{.45\textwidth}
  \centering
  \includeinkscape[inkscapelatex=false,width=\columnwidth]{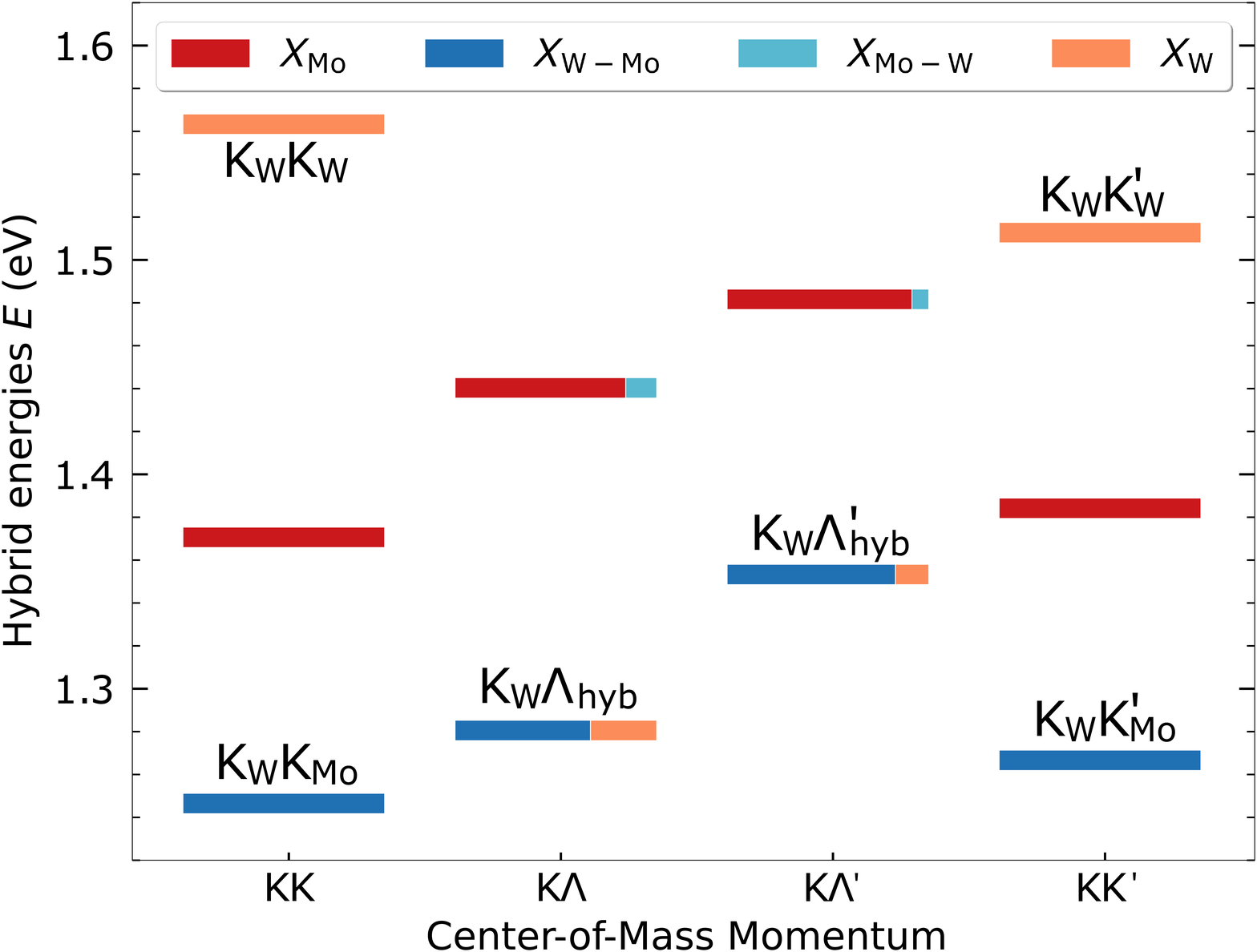}
\end{subfigure}
\hfill
\begin{subfigure}{.45\textwidth}
  \centering
  \includeinkscape[inkscapelatex=false,width=\columnwidth]{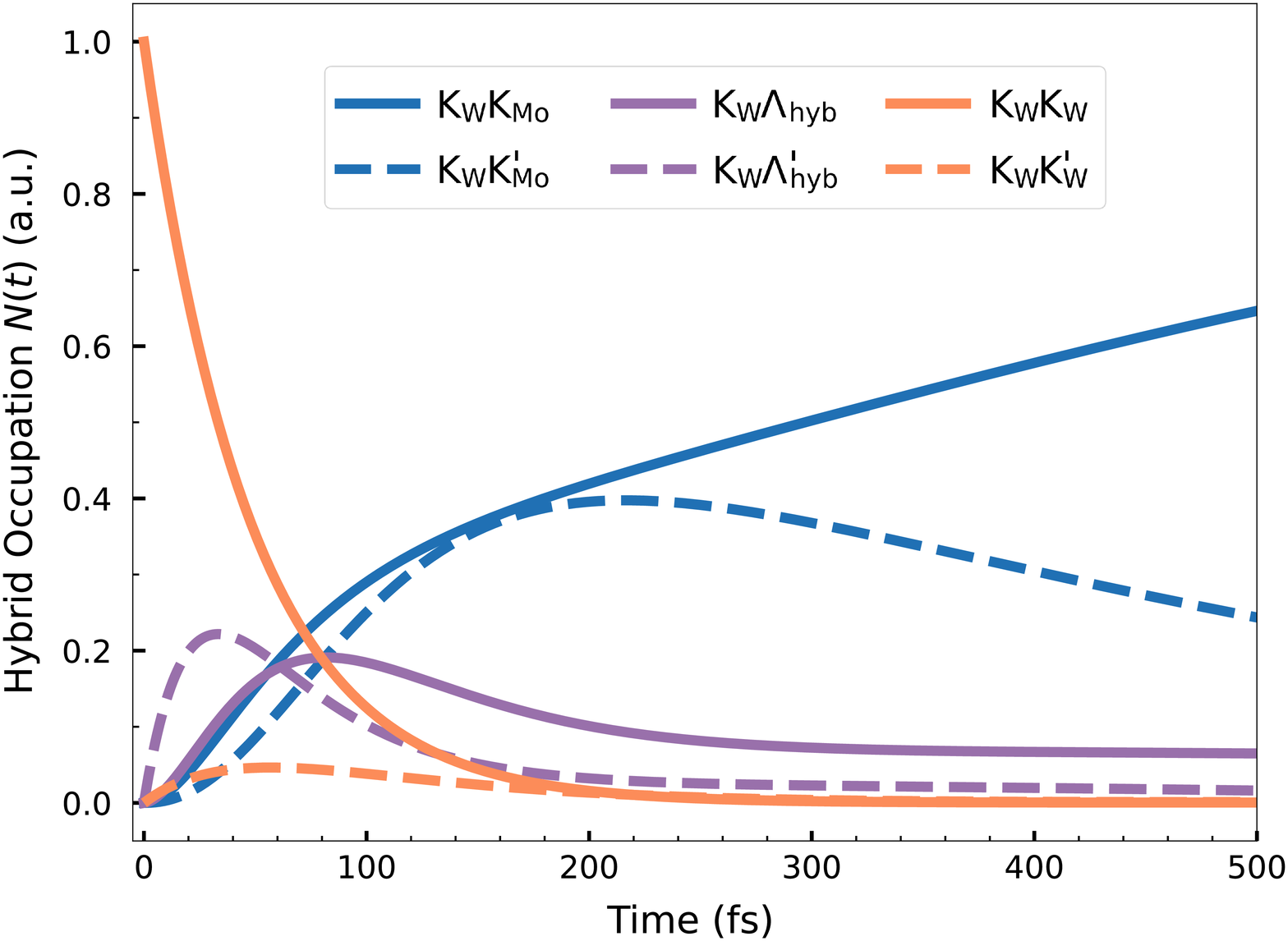}
\end{subfigure}
\caption{(a) Hybrid-exciton energy landscape  and (b) momentum-integrated  dynamics for MoSe$_2$-WSe$_2$ in R$^M_h$ stacking at 77 K. }
\label{RMh}
\end{figure}

\begin{figure}[t!]
\begin{subfigure}{.45\textwidth}
  \centering
  \includeinkscape[inkscapelatex=false,width=\columnwidth]{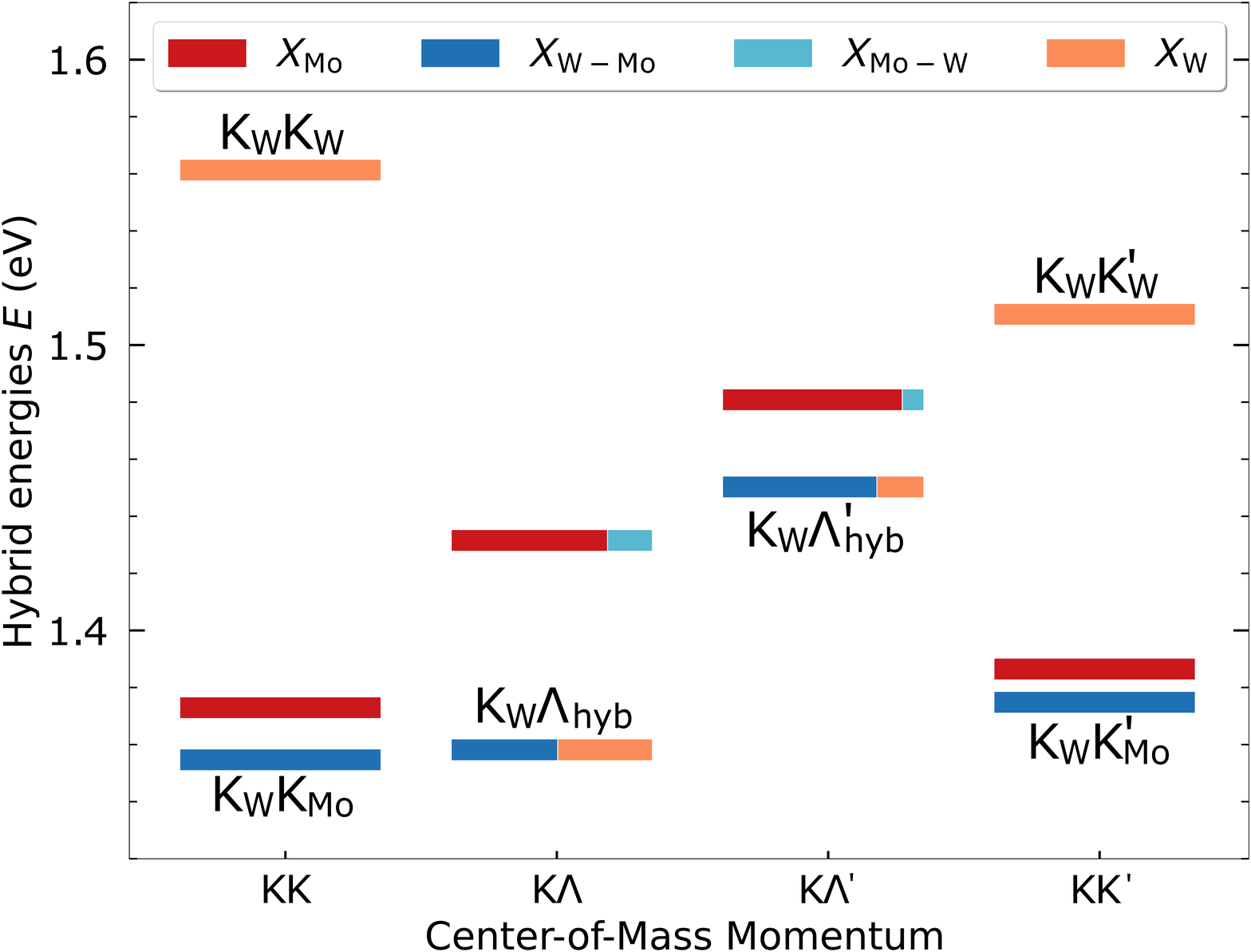}
\end{subfigure}
\hfill
  \begin{subfigure}{.45\textwidth}
  \centering
  \includeinkscape[inkscapelatex=false,width=\columnwidth]{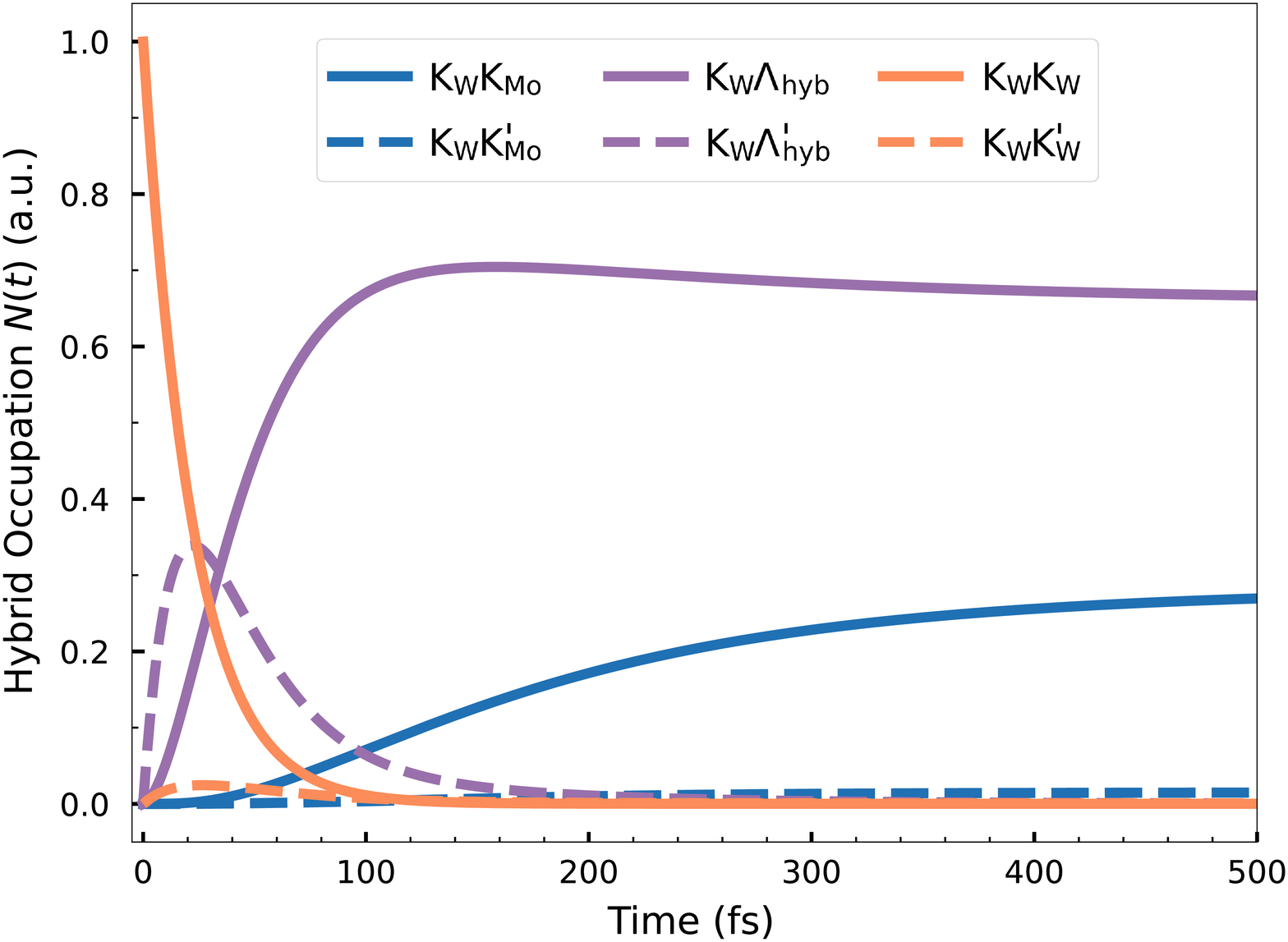}
\end{subfigure}
\caption{(a) Hybrid-exciton energy landscape  and (b) momentum-integrated dynamics for MoSe$_2$-WSe$_2$ in R$^X_h$ stacking at 77 K. }
\label{RXh}
\end{figure}

\section*{Charge transfer in the MoSe$_2$-WSe$_2$ heterostructure}

In the main manuscript, we have discussed the hybrid exciton landscape and dynamics in R$^h_h$ stacking. Here, we show the other two high-symmetry  stackings  R$^M_h$ and R$^X_h$. The main difference originates from a more pronounced tunneling at the $\Lambda$ point. This is due to wave function overlap around the $\Lambda$ point which  has a significantly high contribution also from chalcogen atoms   in these stackings \cite{hagel2021exciton}. This results in larger red-shifts of K$\Lambda_{\text{hyb}}$ excitons bringing them closer to the energetically lowest interlayer K$_\text{W}$K$_\text{Mo}$ excitons, cf. Figs. \ref{RMh}(a) and \ref{RXh}(a).  The hybrid-exciton dynamics remains generally the same as for the R$^h_h$ stacking, i.e. the charge transfer occurs in a two-step process via phonon-mediated scattering into the strongly hybridized K$\Lambda_{\text{hyb}}$ excitons, cf. Figs. \ref{RMh}(b) and \ref{RXh}(b).
The main difference is a larger stationary population of the  hybrid K$\Lambda_{\text{hyb}}$ excitons state as they are closer in energy to the K$_\text{W}$K$_\text{Mo}$ excitons (cf. solid purple line in Figs. \ref{RMh}(b) and \ref{RXh}(b)). This means that the charge transfer is more incomplete compared to the R$^h_h$ stacking. This is in particular the case for the R$^X_h$ stacking, where  
K$\Lambda_{\text{hyb}}$ and K$_\text{W}$K$_\text{Mo}$ excitons are nearly degenerate, cf. Figs. \ref{RXh}(a), as further discussed in the main text.

\begin{figure}[b!]
  \centering
  \includeinkscape[inkscapelatex=false,width=.65\columnwidth]{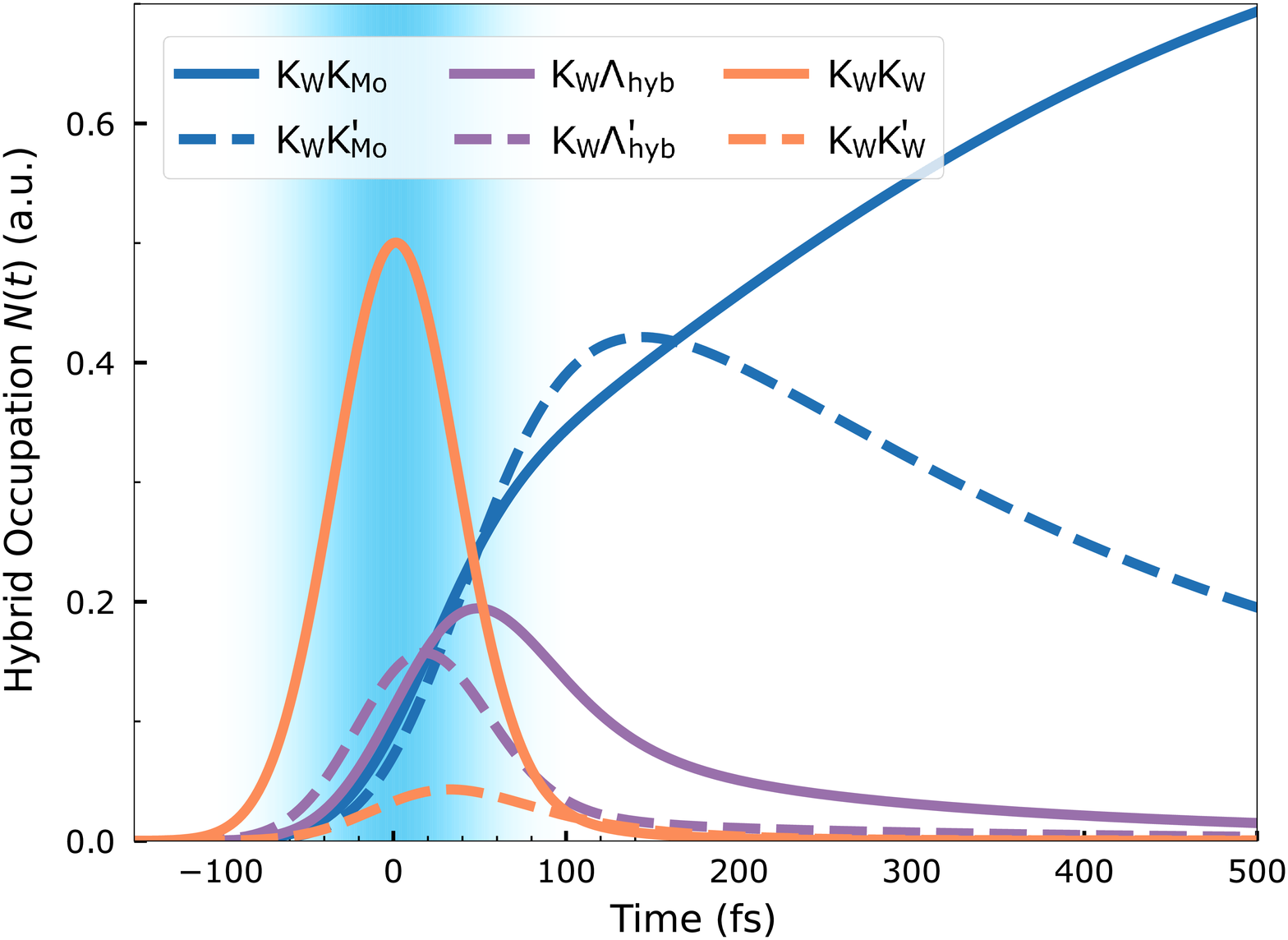}
\caption{Momentum-integrated hybrid-exciton dynamics at 77 K for MoSe$_2$-WSe$_2$ in R$^h_h$ stacking after optical excitation with a laser pulse that is resonant to the K$_\text{W}$K$_\text{W}$ excitons in  the WSe$_2$ layer and has width of 50 fs. During the laser pulse there is an interplay of optical excitation and phonon-mediated hybrid-exciton relaxation and charge transfer processes. }
\label{dyn-pulse}
\end{figure}

\section*{Interplay of optical pumping and dynamics}

While in the main text we have considered the  situation of an instantaneous initial non-equilibrium population in the intralayer K$_\text{W}$K$_\text{W}$ exciton and the subsequent relaxation cascade, we investigate here the hybrid-exciton dynamics taking explicitly into account the interplay of optical pumping and relaxation dynamics. 

We apply a laser pulse with a width of 50 fs and an energy resonant to K$_\text{W}$K$_\text{W}$ excitons and investigate the MoSe$_2$-WSe$_2$ heterostructure in R$^h_h$ at 77 K, cf. Fig. \ref{dyn-pulse}. Comparing the dynamics with the instantaneous initial population in the main text, we observe the same qualitative behaviour in terms of the two-step charge transfer process via phonon-mediated scattering into the strongly hybridized
K$\Lambda_{\text{hyb}}$ excitons. The  main difference occurs in the early stage of the dynamics, i.e.  as soon as  hybrid-excitons are generated they start relaxing to lower energy state. As a result, the maximum occupation of K$_\text{W}$K$_\text{W}$ does not go beyond 0.5. Other than that the hybrid exciton dynamics is almost identical with the one in the main text. However, the interplay of an inward and outward flux of hybrid-excitons from the initial state add an external dependence on the excitation processes, which makes the evaluation of an intrinsic timescale for the charge transfer more complicated. As the dynamics is the same, we have decided to study the charge transfer using the simpler initialization in the main text.

\begin{figure}[t!]
\begin{subfigure}{.44\textwidth}
  \centering
  \includeinkscape[inkscapelatex=false,width=\columnwidth]{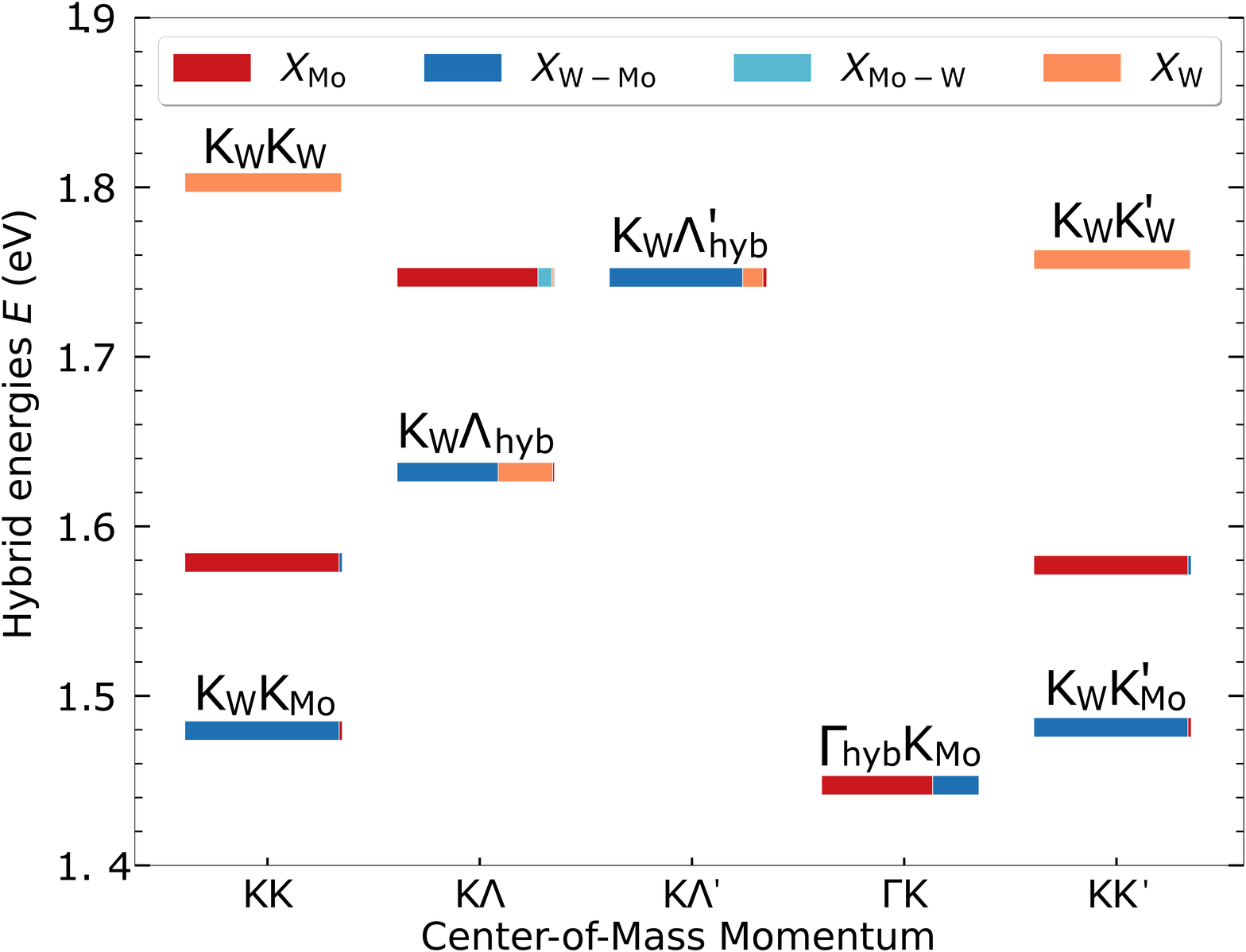}
\end{subfigure}
\hfill
\begin{subfigure}{.44\textwidth}
  \centering
  \includeinkscape[inkscapelatex=false,width=\columnwidth]{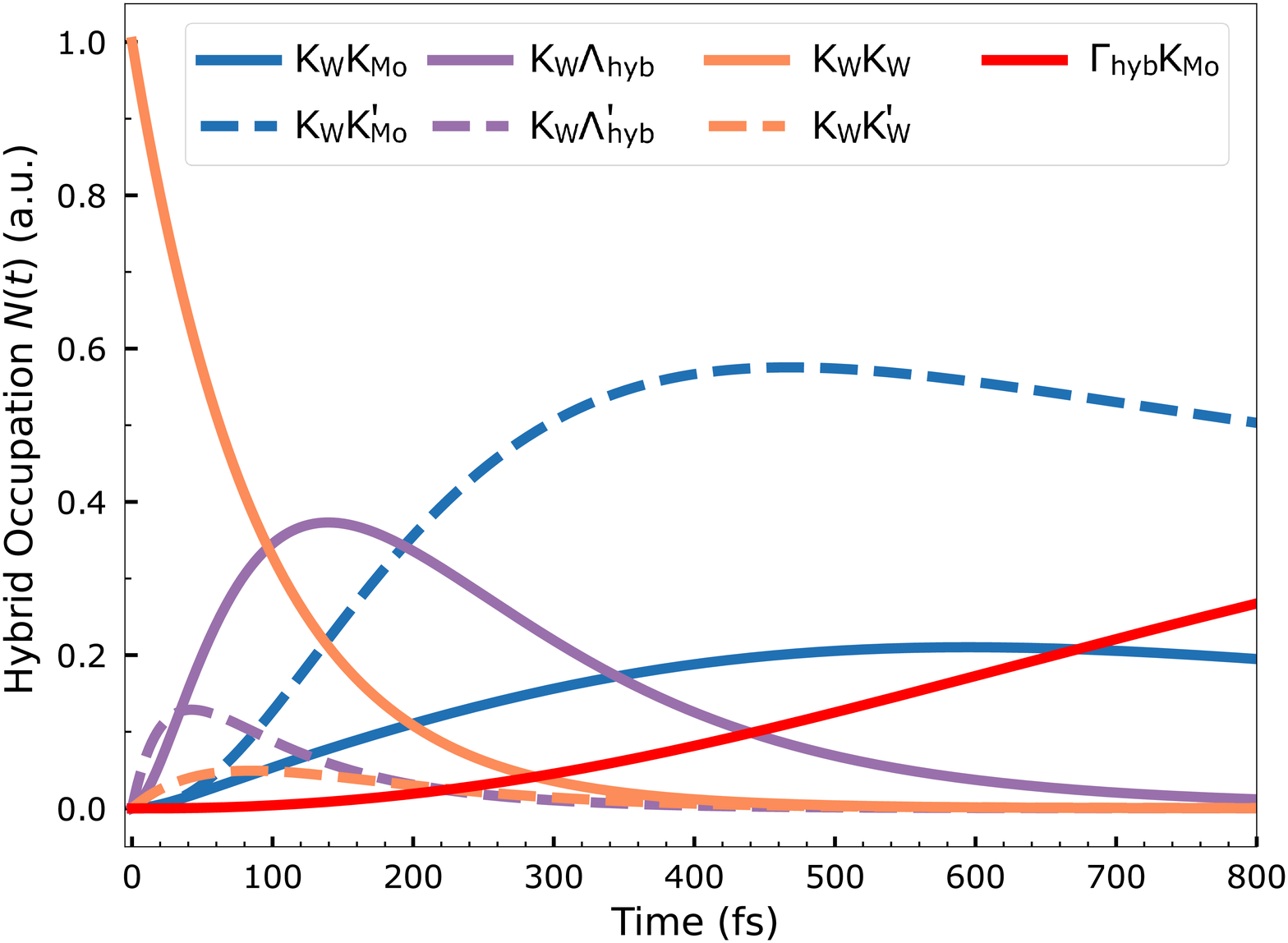}
\end{subfigure}
\caption{(a) Hybrid-exciton energy landscape  and (b) momentum-integrated  dynamics for MoS$_2$-WS$_2$ in R$^h_h$ stacking at 77 K.}
\label{Rhh_MoS2WS2}
\end{figure}

\begin{figure}[t!]
\begin{subfigure}{.44\textwidth}
  \centering
  \includeinkscape[inkscapelatex=false,width=\columnwidth]{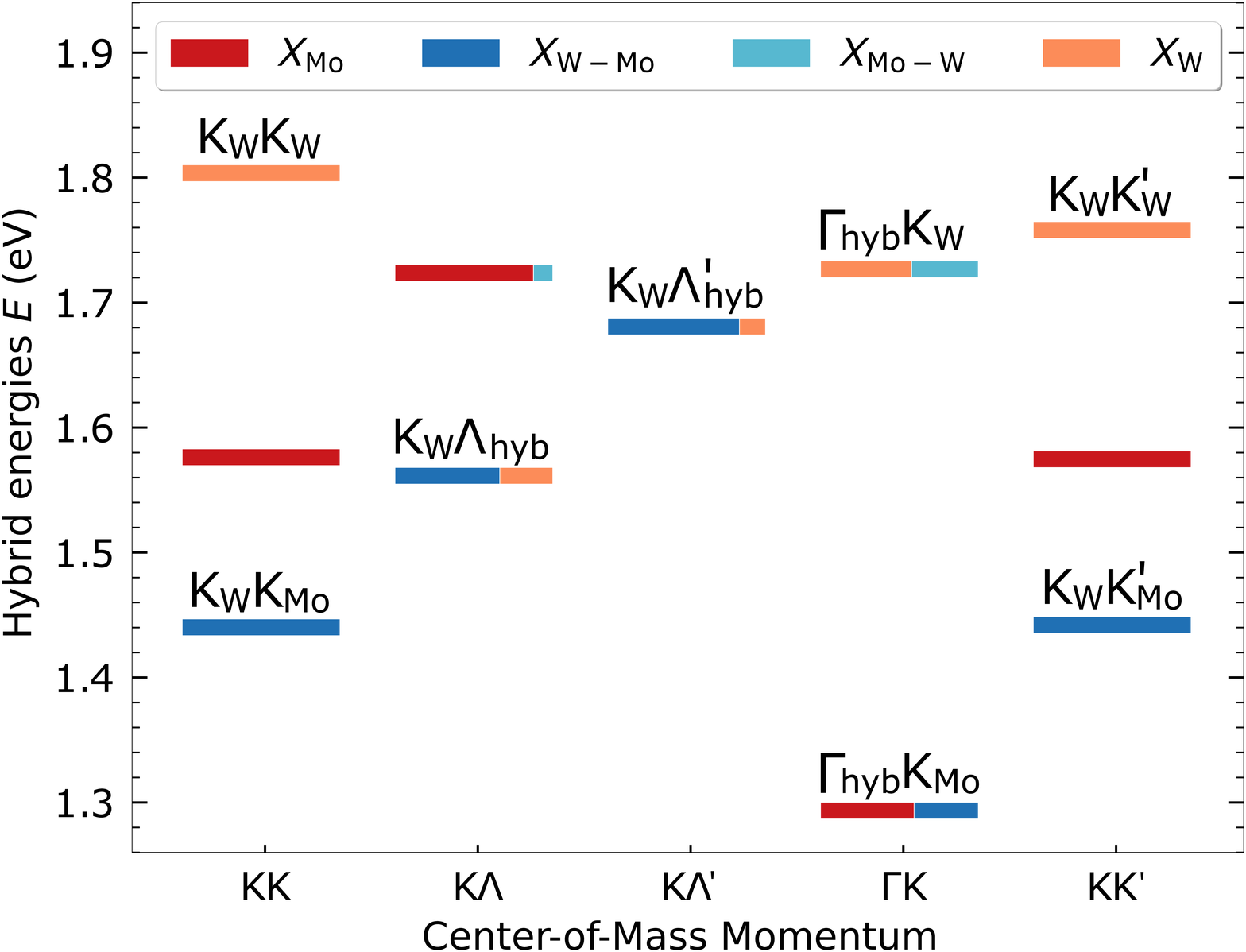}
\end{subfigure}
\hfill
\begin{subfigure}{.44\textwidth}
  \centering
  \includeinkscape[inkscapelatex=false,width=\columnwidth]{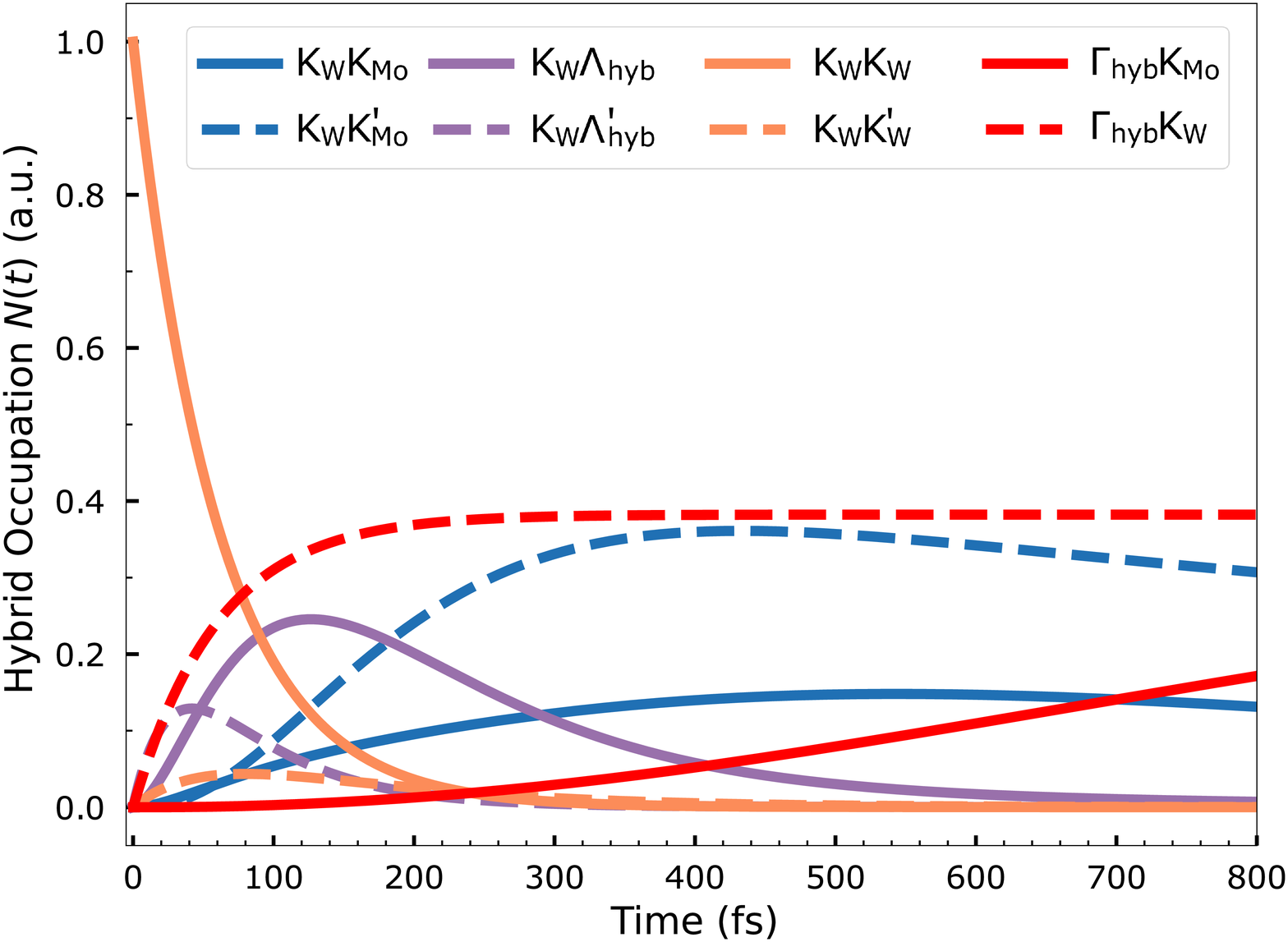}
\end{subfigure}
\caption{(a) Hybrid-exciton energy landscape  and (b) momentum-integrated  dynamics for MoS$_2$-WS$_2$ in R$^M_h$ stacking at 77 K.}
\label{RMh_MoS2WS2}
\end{figure}

\begin{figure}[t!]
\begin{subfigure}{.44\textwidth}
  \centering
  \includeinkscape[inkscapelatex=false,width=\columnwidth]{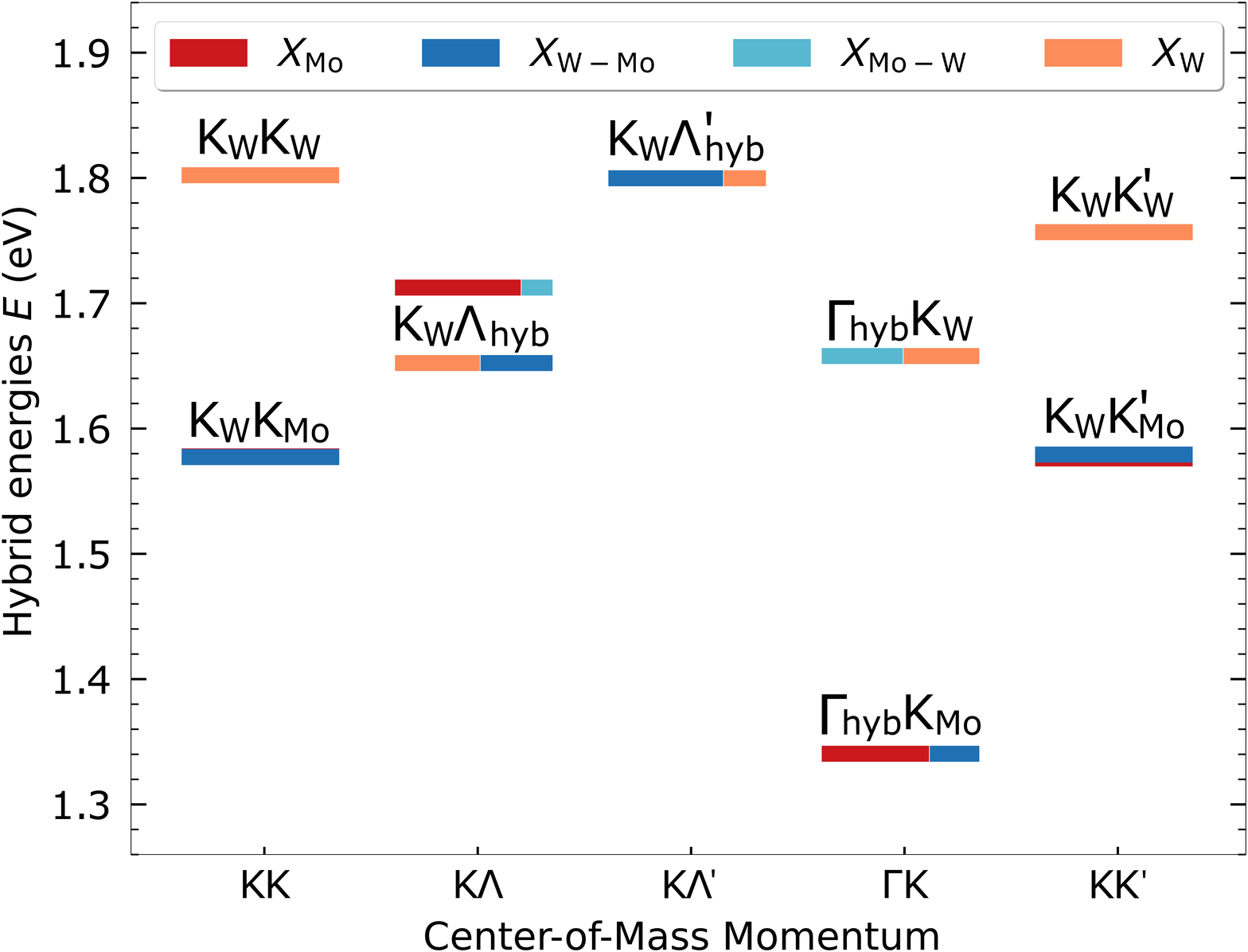}
\end{subfigure}
\hfill
\begin{subfigure}{.44\textwidth}
  \centering
  \includeinkscape[inkscapelatex=false,width=\columnwidth]{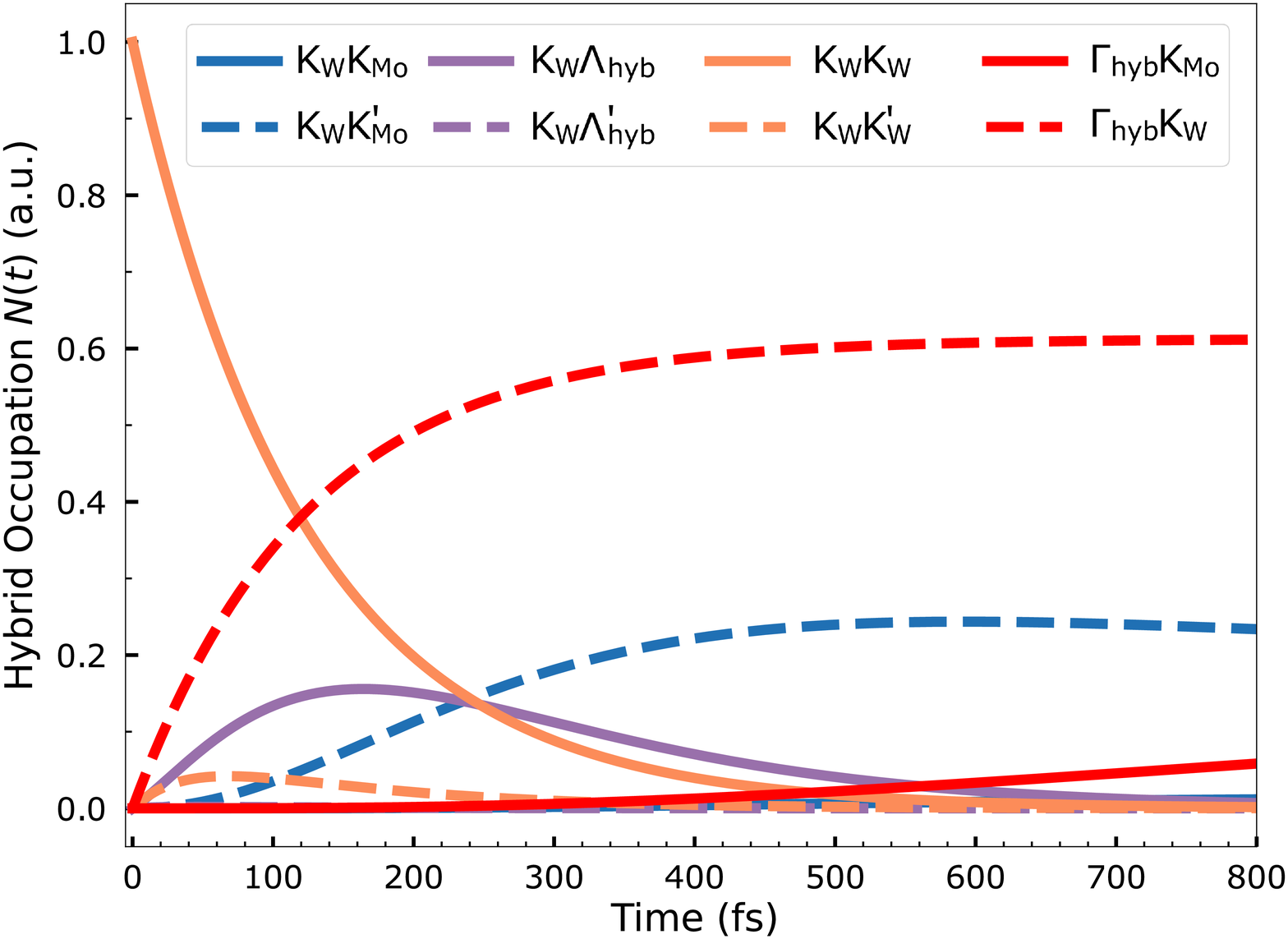}
\end{subfigure}
\caption{(a) Hybrid-exciton energy landscape  and (b) momentum-integrated  dynamics for MoS$_2$-WS$_2$ in R$^X_h$ stacking at 77 K.}
\label{RXh_MoS2WS2}
\end{figure}

\section*{Charge transfer in the MoS$_2$-WS$_2$ heterostructure}

Here, we present the results for the MoS$_2$-WS$_2$ heterostructure and discuss in particular the differences compared to the MoSe$_2$-WSe$_2$ heterostructure discussed in the main part. The main difference in the  hybrid-exciton energy landscape is the appearance of the $\Gamma_\text{hyb}$K$_{\text{Mo}}$ exciton as the energetically lowest state for all three high-symmetry stackings, cf. Figs. \ref{Rhh_MoS2WS2}, \ref{RMh_MoS2WS2}, and \ref{RXh_MoS2WS2}. We find that  it is a highly hybridized state consisting of interlayer and intralayer excitons in the Mo layer. This means that for the 
MoS$_2$-WS$_2$ heterostructure
the charge transfer will be a three-step process. There is first a phonon-mediated electron transfer that occurs via scattering to the hybridized K$_\text{W}\Lambda_\text{hyb}$ excitons to the interlayer K$_\text{W}$K$_\text{Mo}$ states -  similar to the situation in the MoSe$_2$-WSe$_2$ bilayer, discussed in the main part. In contrast to the latter, in MoS$_2$-WS$_2$, the electron transfer is followed by a consecutive hole transfer to the energetically lowest $\Gamma_{\text{hyb}}$K$_{\text{Mo}}$ excitons. This occurs on a slower timescale in the range of hundreds of femtoseconds, cf. the red solid lines in Figs. \ref{Rhh_MoS2WS2}(b), \ref{RMh_MoS2WS2}(b), and \ref{RXh_MoS2WS2}(b).

\begin{figure}[b!]
  \centering
  \includeinkscape[inkscapelatex=false,width=.67\columnwidth]{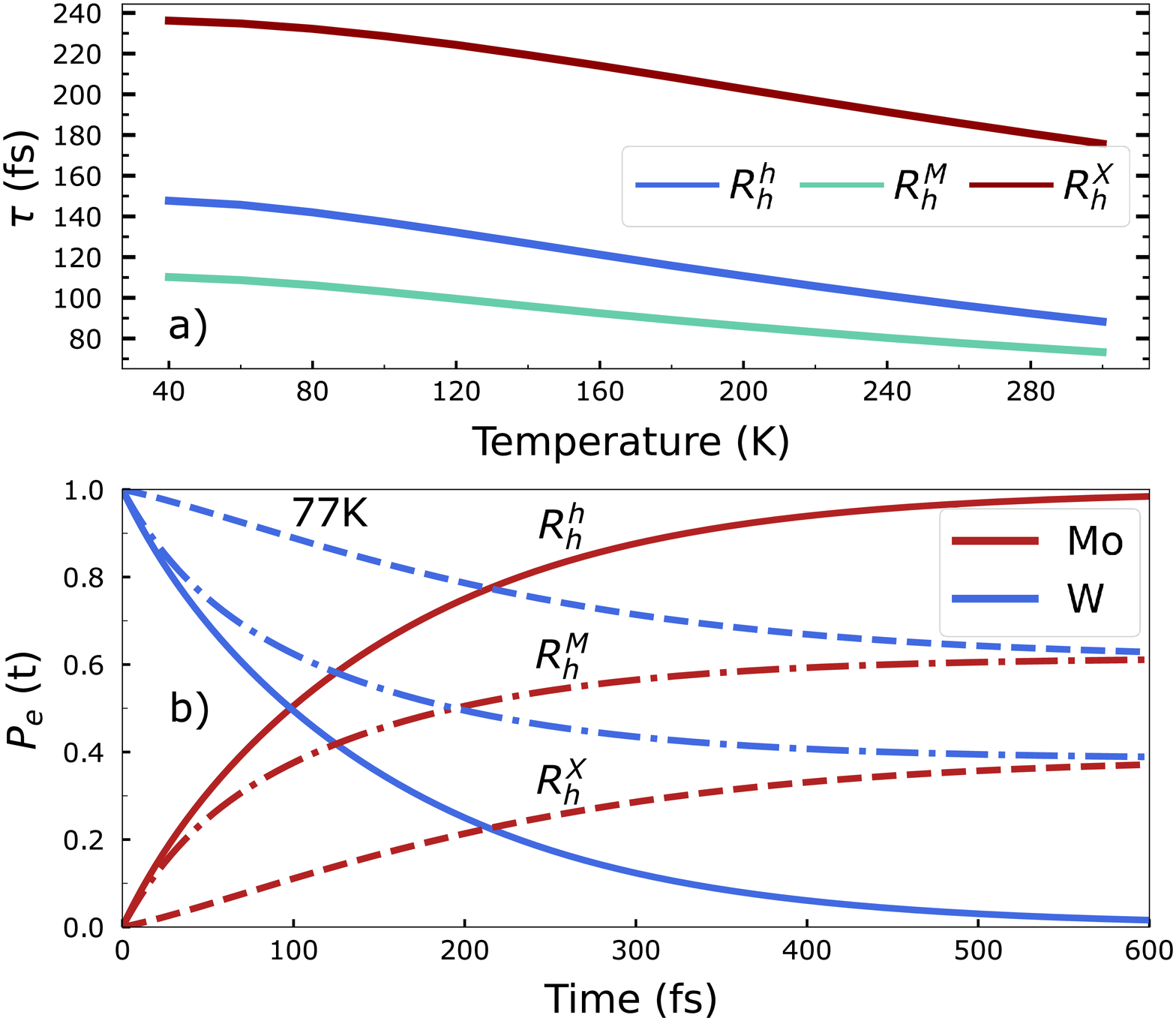}
  \caption{(a) Characteristic charge transfer  time for MoS$_2$-WS$_2$ for all three high-symmetry stackings at 77 K. The time is evaluated through an exponential fit of the electron probability $P_e$ plotted in part (b). We observe a complete charge transfer for $R^h_h$ stacking (solid lines), while there is only a partial charge transfer for $R^X_h$ and $R^M_h$ stackings (dashed and dashed-dotted lines).}
  \label{charge_transfer_MoS2WS2}
\end{figure}

The much larger energy window and the relative energy difference between the states relevant for the relaxation cascade make the charge transfer processes slower in MoS$_2$-WS$_2$. We predict a characteristic time for the charge transfer to be $\tau = 142$ fs for R$^h_h$ stacking at 77 K (Fig. \ref{charge_transfer_MoS2WS2}(a)), which is almost double as large as for MoSe$_2$-WSe$_2$. Furthermore, the stacking dependence is more pronounced, since the strong tunnelling at the $\Gamma$ point and the corresponding red-shift makes the $\Gamma_{\text{hyb}}$K$_\text{W}$ state accessible through scattering with phonons in the case of R$^M_h$ and R$^X_h$ stacking (cf. Figs. \ref{RMh_MoS2WS2}(a) and \ref{RXh_MoS2WS2}(a)). 
This state considerably slows down  the electron transfer, since the population entering the state is trapped and cannot further relax. There is only  the K$_\text{W}\Lambda_\text{hyb}$  state in the vicinity which has a similar composition and could be a scattering partner. However, the scattering into this state  requires a simultaneous electron and hole scatter, which is a weak high-order process.
As the $\Gamma_{\text{hyb}}$K$_\text{W}$ state is relatively low in energy  in the R$^X_h$ stacking, the percentage of population trapped in this state is higher with respect to the R$^M_h$ stacking ($P_e = 0.38$ for R$^X_h$ against $P_e = 0.61$ for R$^M_h$, cf. Fig. \ref{charge_transfer_MoS2WS2}(b)). This explains the slowest charge transfer found for  R$^X_h$ stacking.

\bibliography{references}